\documentstyle[epsf,epsfig,rotate]{mn}
\newcommand{\be}{\begin{equation}}
\newcommand{\ee}{\end{equation}}

\newcommand{\apj}{{\it ApJ, }}

\newcommand{\icar}{{\it Icarus, }}

\newcommand{\ana}{{\it A\&A, }}

\title
[MHD turbulence and protoplanets]
{The interaction of planets with a disc with MHD turbulence IV:
Migration rates of embedded protoplanets}
\author[R.P.Nelson \& J.C.B.Papaloizou]{Richard P. Nelson \&
John C.B. Papaloizou \\
Astronomy Unit, Queen Mary, University of London, Mile End Rd, London E1 4NS}

\date{Received/Accepted}

\begin{document}

\maketitle

\begin{abstract} 
We present the results of global cylindrical disc simulations
and local shearing box simulations of protoplanets interacting
with a disc undergoing MHD turbulence. The specific emphasis
of this paper is to examine and quantify the magnitude of the
torque exerted by the disc on the embedded protoplanets as a function
of the protoplanet mass, and thus to make a first
study of  the induced  orbital migration of protoplanets resulting from their
interaction with 
magnetic, turbulent discs. This issue is of crucial importance in
understanding
the formation of gas giant planets through the so--called core instability
model, and the subsequent orbital evolution post formation prior
to the dispersal of the protostellar disc. Current estimates
of the migration time of protoplanetary cores in the $3$ -- $30$
Earth mass range  in standard disc models
are $\tau_{mig} \simeq 10^4$ --  $10^5$ yr, which is much shorter than
the estimated gas accretion time scale of Jupiter type planets. 

The global simulations were  carried out for a disc with aspect ratio
$H/R =0.07$
and protoplanet masses of $M_p= 3, 10$, $30$ Earth masses,
and 3 Jupiter masses.
The local shearing box simulations were carried out for
values of the dimensionless parameter $(M_p/M_*)/ (H / R)^3
=0.1, 0.3, 1.0, $ and $2.0,$ with $M_*$ being the central mass.
These allow both embedded and gap forming
protoplanets for which the disc response is non linear to be investigated.

In all cases the instantaneous net torque experienced by a protoplanet
showed strong fluctuations on an orbital time scale,
and in the low mass embedded cases
oscillated between negative and positive values. Consequently,
in contrast to the laminar disc type I migration scenario,
orbital migration would occur as a random walk.
Running time averages for embedded protoplanets 
over typical run times of  $20$ -- $25$ orbital periods, 
indicated that the averaged torques from the inner and outer disc
took on values characteristic of
type I migration. However, large fluctuations occurring on longer than 
orbital time scales remained,
preventing convergence  of the average torque to well defined
values or  even to a well defined sign for these lower mass cases.

Fluctuations became relatively
smaller for larger masses indicating better convergence properties, to the
extent that
in the $30 M_{\oplus}$ simulation consistently inward, 
albeit noisy, migration was indicated.

Both the global and local simulations showed this
trend  with  increasing
protoplanet mass which is due to its perturbation on the disc
increasing to become comparable to and then  dominate the turbulence in its neighbourhood.
This then becomes unable to produce very large long term  fluctuations in the
torques acting on the protoplanet. 
Eventually gap formation
occurs and there is a transition to the usual 
type II
migration at a
rate determined by the angular momentum transport in the distant parts of the  disc.

The existence of significant fluctuations occurring
in the turbulent discs on long time scales is an important unexplored issue
for the lower mass embedded protoplanets, that are unable
to modify the turbulence in their neighbourhood,
and which have been  studied here.
If  significant fluctuations 
occur on the longest disc  evolutionary time scales, convergence
of torque running averages for practical purposes  will not occur
and
the migrational behaviour of low mass protoplanets considered as an ensemble
would be very different from predictions of type I migration theory
for laminar discs.
The fact that  noise levels were relatively smaller in the local  simulations
may indicate the presence of long term global fluctuations, but the issue remains an
important one for future investigation.

\end{abstract}


\section{Introduction}\label{S0} 
\noindent

The ongoing discovery of extrasolar giant planets has stimulated
renewed interest in the theory of planet formation 
(e.g. Mayor \& Queloz 1995;
Marcy, Cochran, \& Mayor 1999; Vogt et al. 2002; Santos et al. 2003).
In the most commonly accepted theory of how planets form, the so--called
core instability model, gas giant planets form through the 
build--up of a rocky and icy core of $\sim 15$ Earth masses, which then
undergoes gas accretion resulting in a gas--giant planet
(e.g. Bodenheimer \& Pollack 1986;
Pollack et al. 1996). An alternative model involves the formation of
giant planets through the gravitational fragmentation of the protostellar disc
during the earlier phases of its evolution (Boss 2001). In either case
disc--planet interaction will play an important role in the subsequent
evolution.

The gravitational interaction between protostellar discs and embedded
protoplanets has been the subject of a large number of
studies over the last couple of decades.
In the standard picture, a protoplanet exerts torques on a protostellar disc
through the excitation of spiral density waves at Lindblad resonances,
and possibly through interaction at corotation resonance
(e.g. Goldreich \& Tremaine 1979; Lin \& Papaloizou 1979;
Papaloizou \& Lin 1984; Ward 1986, 1997;
Tanaka, Tacheuchi \& Ward 2002).
The spiral waves carry with them an associated angular momentum flux.
This angular momentum is deposited in the disc material
where the 
waves are damped, leading to an exchange of angular momentum
between protoplanet and disc. 
The disc that lies exterior to the protoplanet orbit exerts a negative torque on
the planet, and the interior disc exerts a positive torque.
For most disc models the negative torque dominates and the protoplanet
migrates inwards. For protoplanets of $\sim 15$ Earth masses, the migration
time is estimated to be between $10^4$ and $10^5$ yr (Tanaka, Tacheuchi, \& 
Ward 2002), which is very much
shorter than the estimated gas accretion phase of giant
planet formation $\simeq 7$ Myr (Pollack et al. 1996).  Taken at face value,
this presents a serious problem for the core--instability model of
gas giant planet formation. We note, however, this analysis pertains only to
smooth, laminar disc models.

For protoplanets in the Jovian mass range, the interaction is non linear
and gap formation occurs 
(Papaloizou \& Lin 1984; Bryden et al. 1999; Kley 1999). In this case the
orbital migration of the planet becomes locked to the viscous evolution of
the disc, and migration is expected to occur on a time scale
of $10^5$ yr (Lin \& Papaloizou 1986; Nelson et al. 2000; D'Angelo, Kley \&
Henning 2002).

Until quite recently most models of viscous accretion discs used the
Shakura \& Sunyaev (1973) $\alpha$ model for the anomalous
disc viscosity. This assumes
that the viscous stress is proportional to the thermal pressure in the disc,
without specifying the origin of the viscous stress (but assumed to arise from
some form of turbulence). Work by Balbus \& Hawley (1991) indicated that
significant angular momentum transport in weakly magnetised discs could arise 
from the magnetorotational instability (MRI -- or the Balbus--Hawley 
instability).
Subsequent non linear numerical simulations performed using a local
shearing box formalism
(e.g. Hawley \& Balbus 1991; 
Hawley, Gammie, \& Balbus 1996; Brandenburg et al. 1996)  confirmed
this and showed that the saturated non linear outcome of the MRI
is MHD turbulence
with an associated viscous stress parameter $\alpha$ of between $\sim 5 \times
10^{-3}$ and $\sim 0.1$, depending on the initial magnetic field configuration.
More recent global simulations of MHD turbulent discs
[e.g. Armitage 1998; Hawley 2000; 
Hawley 2001; Steinacker \& Papaloizou 2002; Papaloizou \& Nelson 2003] 
confirm the picture 
provided by the local shearing box simulations. 

This is the fourth in a series of papers that examine the interaction between
disc models undergoing MHD turbulence with zero net flux magnetic
fields and embedded protoplanets. In Papaloizou \& Nelson (2003 -- hereafter
paper I)
we examined and characterised the turbulence obtained in a variety
of MHD disc models. In Nelson \& Papaloizou (2003 -- hereafter paper II) we examined the
interaction between a global cylindrical disc model and a massive (5 Jupiter
mass) protoplanet. A similar study was undertaken by 
Winters, Balbus, \& Hawley (2003b). In a companion paper to this one
(Papaloizou, Nelson, \& Snellgrove 2003 -- hereafter paper III)
we presented the results of global cylindrical disc simulations
and local shearing box simulations of turbulent discs interacting with
protoplanets of different mass. The main focus of paper III was to
characterise the changes in flow morphology and disc structure as
a function of planet mass, and to examine the transition from
linear to non linear interaction leading to gap formation. In this paper
we continue to examine these simulations, but now focus on
the gravitational torques exerted on the protoplanet by the disc
and the associated migration rate of the protoplanet.

We find that in all simulations performed, 
the torque experienced by the protoplanet
is a highly variable quantity on account of the
protoplanet interacting with the turbulent density wakes that shear past it.
For low mass protoplanets, the torque is dominated by
these fluctuations, such that the usual distinction between inner (positive)
and outer (negative)
disc torques is blurred. The net torque experienced by embedded
protoplanets oscillates between negative and positive values, such that
the protoplanet migration is likely to occur as a random walk.
This is in contrast to the monotonic inward drift normally associated with 
type I migration. A running time average of the torques fails to converge 
for the embedded protoplanet runs, at least for the run times that are currently
feasible, so that definitive statements
about the direction and rate of migration of low mass planets in turbulent
discs cannot yet be made. 

In a manner that is consistent with the results
of paper III, we find that the results show a definite trend as a function 
of planet mass. For very low mass planets the turbulent density wakes 
are of much higher amplitude than the spiral wakes generated by the
planet. This is reflected in the torque experience by the planet,
for which the turbulent fluctuations are very much larger than the running mean
torque. As the planet mass increases, the spiral wakes generated by the planet
become apparent in the flow. This is accompanied by the torques
displaying the expected separation between inner and outer disc torques,
and the fluctuations becoming smaller relative to the running mean 
torque.
This trend with increasing mass continues until gap formation
occurs. At this point  we find a weakening  of the torques
exerted by the disc on the protoplanet due to the evacuation of gap material.
There is then a transition to type II
migration and,  as exemplified  by  the simulation of a $5$ Jupiter mass
protoplanet interacting with a turbulent disc presented in paper II,
inward migration at a
rate determined by the angular momentum transport in the distant parts of the  disc
unaffected by the protoplanet.

The plan of the paper is as follows. In section~\ref{S1} 
we described our initial
model set up and numerical procedure. In section~\ref{S3a} we
 present the results of the simulations.
Finally in section~\ref{conclusions} we discuss our results.

\section{Initial model setup} \label{S1} 
The initial set up and boundary conditions used in the models presented
in this paper are described in detail in paper III, and we do not
repeat that discussion here. However, we present some details of the
runs in tables~\ref{table1} and \ref{table2} for convenience.
Table~\ref{table1} contains details of the global runs, and table~\ref{table2}
contains details of the local shearing box runs.

\subsection{Global Runs}

The initial conditions for the global runs consisted of a turbulent
accretion disc model with $H/r=0.07,$ $H$ being the putative disc semi
thickness, and  a volume averaged characteristic
Shakura--Sunyaev $\alpha$ value of $\alpha \simeq 7 \times 10^{-3}$,
as described in paper III. At time
$t=0$ the protoplanet was inserted into the disc and subsequently
held fixed  at location 
$(r_p,\phi_p)=(3,\pi)$ in cylindrical
coordinates $(r, \phi)$ centred on the central mass $M_*.$ 
The gravitational potential of the protoplanet
was softened using a softening parameter $b=0.3H_p$, where $H_p$ is the
disc scale height at the planet location.
The simulation was performed in a rotating
reference frame whose angular velocity equalled the Keplerian
angular velocity of the protoplanet. The torque on the planet as a 
function of time was calculated in the manner described in section~\ref{torque}.
The unit of time used when discussing the global runs is the orbital period
of the protoplanet $P_p=2 \pi/\Omega_p$.

\begin{table*}
 \begin{center}
 \begin{tabular}{|l|l|l|l|l|l|l|l|}\hline\hline
       &     &     &     &     &        &        &              \\
 Model&$\phi$ domain&$H/r$&$M_p/M_*$& $(M_p R^3/(M_* H^3)$&$n_r$&$n_{\phi}$&$n_z$\\
       &     &     &     &     &        &        &
  \\
 \hline
 \hline
G1 & $2 \pi$ & 0.07 & $1 \times 10^{-5}$ & 0.03 & 450 & 1092 & 40 \\
G2 & $2 \pi$ & 0.07 & $3 \times 10^{-5}$ & 0.09 & 450 & 1092 & 40 \\
G3 & $2 \pi$ & 0.07 & $1 \times 10^{-4}$ & 0.30 & 450 & 1092 & 40 \\
G4 & $2 \pi$ & 0.07 & $3 \times 10^{-3}$ & 8.75 & 450 & 1092 & 40 \\
G5 & $\pi/2$ & 0.07 & $3 \times 10^{-3}$ & 8.75 & 450 & 276 & 40 \\
\hline
\end{tabular}
\end{center}
\caption{ \label{table1}
Parameters of the global simulations: The first column gives the simulation
label, the second gives the extent of the azimuthal domain, the third gives
the $H/r$ value of the disc, and the fourth gives the
protoplanet-star mas ratio. The fifth gives the ratio of $M_p/M_*$ to
$(H/r)^3$. The sixth, seventh, and eighth columns describe the number of
grid cells used in each coordinate direction.}
\end{table*}

\subsection{Shearing Box Runs} \label{sh_box}

\begin{table*}
 \begin{center}
 \begin{tabular}{|l|l|l|l|l|l|l|l|l|l|l|}\hline\hline
       &     &     &     &     &        &        &        &          \\
 Model&$z$ domain&$x$ domain&$y$ domain &$t_1$&$t_2$&$(M_p R^3/(M_* H^3)$&$n_z$&$n_x$&$n_y$\\
       &     &     &     &     &        &        &        &     &     &
  \\
 \hline
 \hline
 Ba1&$(-H/2, H/2)$& $(-4H, 4H)$ & $(-2\pi H, 2\pi H)$ & $354$&  $621$&$0.1$&$35$&$261$&$200$\\
 Ba2&$(-H/2, H/2)$& $(-4H, 4H)$ & $(-2\pi H, 2\pi H)$ & $354$&  $644$&$0.3$&$35$&$261$&$200$\\
 Ba3&$(-H/2, H/2)$& $(-4H, 4H)$ & $(-2\pi H, 2\pi H)$ & $354$&  $619$&$1.0$&$35$&$261$&$200$\\
 Ba4&$(-H/2, H/2)$& $(-4H, 4H)$ & $(-2\pi H, 2\pi H)$ & $354$&  $650$&$2.0$&$35$&$261$&$200$\\
\hline
\end{tabular}
\end{center}
\caption{ \label{table2}
Parameters of the shearing box 
simulations: The first column gives the simulation label,
the second, third and fourth
give the extent of the coordinate domains considered.
The $x,y,$ and $z$ domains refer to the Cartesian domains labeled
as radial, azimuthal and vertical respectively.
The fifth and sixth column
give the start and end times if the simulation measured in dimensionless units
described in section~\ref{sh_box}. 
These were all started by inserting the protoplanet into  the turbulent
model generated from the simulation Ba0 described in paper III  at time indicated.
The seventh column
gives the value of $(GM_p/(\Omega_p^2 H^3) = (M_p R^3/(M_*  H^3).$
 The eighth, ninth and tenth columns
give the number of computational grid points in the $x,y,$ and $z$ coordinates respectively.
These numbers include any ghost zones used to handle boundary conditions.}
\end{table*}

The morphology of the shearing box runs has already been
described in detail in paper III. These were generated from
a simulation Ba0 described there which had fully developed turbulence, 
but no protoplanet, that had been run for 354 time units. 
The time unit  was taken to be the inverse angular velocity ,$\Omega_p,$ 
 at the centre
of the box located a putative distance $R$
from the central object. Simulations Ba1--Ba4 were then carried out after inserting protoplanets
with a range of values for $GM_p /( H^3\Omega_p^2) = M_p R^3/(M_* H^3),$
with $M_p$ being the protoplanet mass.
These and other  relevant 
parameters are given in table~\ref{table2}.
As for the global simulations, the protoplanet potential
was softened, by use of a softening parameter
$b = 0.3H.$  For additional details see paper III.
The results of these simulations are presented in section \ref{box-res}.

\subsection{Numerical procedure} \label{S2}
\noindent
The numerical scheme that we employ is
based on a spatially second--order accurate method that computes the
advection using the monotonic transport algorithm (Van Leer 1977).
The MHD section of the code uses the
method of characteristics
constrained transport (MOCCT) as outlined in Hawley \& Stone (1995)
and implemented in the ZEUS code.
The code has been developed from a version
of NIRVANA originally written by U. Ziegler (Ziegler \& Rudiger 2000).

\subsection{Torque Calculation} \label{torque}
The torque experienced by the protoplanet in the global disc runs
was calculated by summing
the gravitational force due to the mass in each grid cell in the active domain
of the disc model. Those cells that lay within the boundary layers located near
the radial boundaries of the computational domain
(described in paper III) did not contribute to the gravitational force.
The gravitational force due to the protoplanet acting on the disc
was softened using $b=0.3H_p$ where $H_p$ is the  putative scale height at the
position of the protoplanet. An identical softening was used when calculating 
the force of the disc on the protoplanet. Material that lay inside
the Hill sphere of the protoplanet given by $R_H=r_p(M_p/3M_*)^{1/3}$
was excluded from the torque calculation. The acceleration experienced by 
the protoplanet can be written as
$$ {\bf f} = \sum_{i=1}^{N_{cells}} - 
\frac{G \; m_i \; ({\bf r}_p - {\bf r}_i)}
{(r_p^2 + r_i^2 - 2 r_p r_i \cos{(\phi_p-\phi_i)} +b^2)^{3/2}}$$
where the sum is over  all possible values of the subscript  $i$
which is used to label quantities evaluated at the centre
of  a specific grid cell, thus  ensuring that the whole of the
computational domain is covered.
The mass in each grid cell $m_i = \rho_i \, \Delta r \, r_i \, 
\Delta \phi \, \Delta z$, where $\Delta r$,
$\Delta \phi$, and $\Delta z$ are the grid spacings in the radial, azimuthal,
and vertical directions respectively.
The resulting torque per unit mass is then ${\bf T}= {\bf r}_p \times {\bf f}$.
This quantity was calculated and stored every 10 time steps.

A similar procedure was used to calculate the force acting on the planet
in the shearing box simulations.

\section{Numerical results} \label{S3a}
Before discussing the results of the global turbulent disc simulations, we
present the results of some simulations designed to calibrate the
code. In particular, we will examine the role that gravitational
softening of the protoplanet gravitational potential has on the
migration time in laminar disc simulations.
 
\subsection{Type I migration in laminar discs} \label{calibration}
The migration time of a non gap forming protoplanet embedded in a three
dimensional disc has been calculated using a linear
analysis by Tanaka, Takeuchi, \& Ward (2002). They derive the following 
expression for the migration time of a protoplanet embedded in a
locally isothermal  disc
with a power law surface density profile:

\begin{equation}
\tau_{mig}=(2.7 + 1.1 \gamma)^{-1} \frac{M_*}{M_p} \frac{M_{*}}{{\Sigma_p r^2_p}}
\left(\frac{c}{r_p \Omega_p}\right)^2 \Omega_p^{-1} \label{ward-tmig} 
\end{equation} 
Here $\Sigma_p$ is the surface density at the position of the protoplanet,
$\Omega_p$ is the Keplerian angular velocity of the protoplanet, $c$ is the
sound speed in the disc at the protoplanet position, and
$- \gamma$ is the power law index for the radial surface density distribution.
These authors comment
that in general this expression gives a migration time that is a factor
of between 2 -- 3 times slower than similar expressions derived for
flat, two--dimensional discs (e.g. Ward 1997).
The finite disc thickness thus acts to soften the disc--planet
gravitational interaction. 

\begin{figure}
\centerline{
\epsfig{file=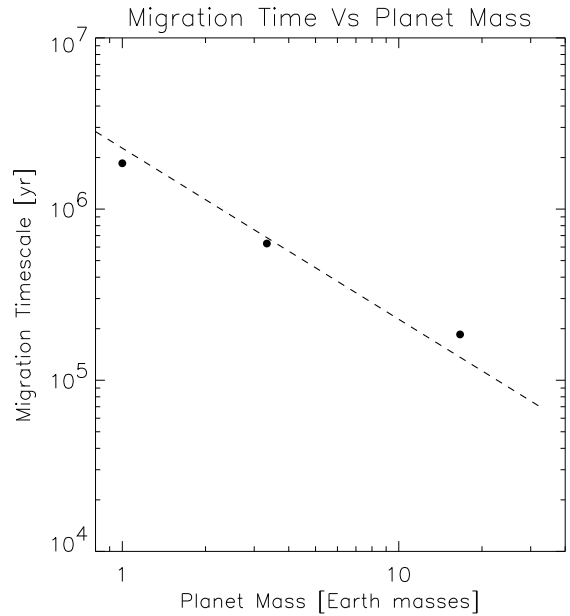,width=\columnwidth} }
\caption[]
{This figure shows the migration time for protoplanets of different mass
using a gravitational softening parameter $b=0.7H_p$. The dots show the
results of the simulations, and the dashed line shows the migration time
computed from equation~\ref{ward-tmig}.}
\label{fig1}
\end{figure}

\begin{figure}
\centerline{
\epsfig{file=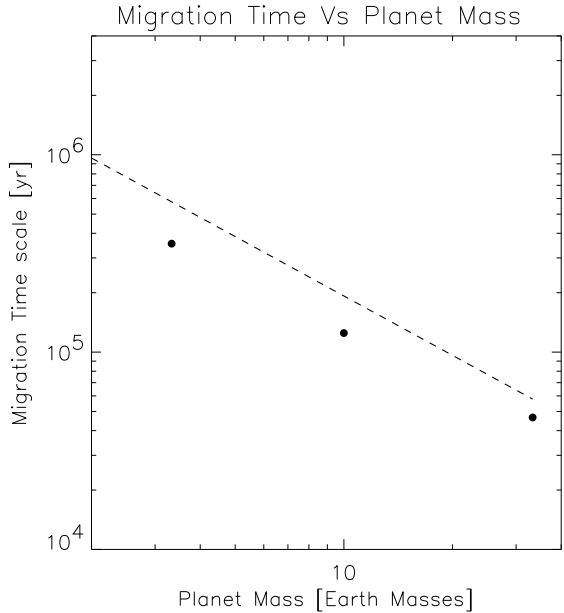,width=\columnwidth} }
\caption[]
{This figure shows the migration time for protoplanets of different mass
using a gravitational softening parameter $b=0.3H_p$. The dots show the
results of the simulations, and the dashed line shows the migration time
computed from equation~\ref{ward-tmig}.}
\label{fig2}
\end{figure}

The global turbulent disc simulations presented here 
use cylindrical disc models in which the
vertical component of gravity is neglected, and thus the full three dimensional
structure of the disc is not modeled. In the limit of a laminar disc, the
cylindrical disc can be viewed as a series of two dimensional
discs stacked on top of one another, providing the protoplanet gravitational
potential is also cylindrical (i.e. no vertical component). Given that 
gravitational softening of the protoplanet potential is employed, we are 
interested in ({\it i}) How large a gravitational softening is required
in a two dimensional disc simulation to obtain agreement with
equation~\ref{ward-tmig}; ({\it ii}) What is the deviation from 
equation~\ref{ward-tmig} obtained for differing softening parameters.
In order to address these points a number of two dimensional
simulations using laminar, viscous $\alpha$--discs 
have been performed with varying physical and 
softening parameters.

Figure~\ref{fig1} shows the migration time for a series of
models in which protoplanets of differing masses were placed in
a protostellar disc model and the torque of the disc acting on the
protoplanet was calculated. These torques were then used to calculate 
a migration time using the expression
\begin{equation}
\tau_{mig} = \frac{r_p}{{\dot r_p}} = \frac{1}{2} \frac{j_p}{T_p} 
\label{tmig-sim} 
\end{equation}
where $j_p$ is the specific angular momentum of the protoplanet and
$T_p$ is the torque per unit mass due to the disc.
The disc model used for the particular set of simulations presented
in figure~\ref{fig1} was such that the disc surface density
$\Sigma(r) = \Sigma_0 r^{-1/2}$, $H/r=0.05$, and the dimensionless viscosity
coefficient $\alpha = 4 \times 10^{-3}$.
In code units the inner boundary was located at $R_{in}=1$,
the outer boundary at $R_{out}=8$, and the protoplanet was located at
$r_p=3$. If we use the convention that $r_p$ is equivalent to 5.2 AU,
and the central stellar mass is a solar mass, then the disc
mass was normalised so that it contained the equivalent of 7.5 Jupiter
masses between 1.56 and 20.8 AU. This disc model is equivalent
to that described in Bate et al. (2003), and figure~\ref{fig1} is directly
comparable with their figure~10. The gravitational softening parameter
was set to be $b=0.7H_p$ where $H_p$ is the disc thickness at the 
position of the protoplanet. We use the convention that 1 Earth mass corresponds
to $M_p/M_*=3 \times 10^{-6}$. 

The black dots in figure~\ref{fig1} show the migration time scales
obtained in the simulation. The dashed line is obtained from
equation~\ref{ward-tmig} assuming an identical disc model.
It is clear that a gravitational softening parameter of
$b=0.7H_p$ gives  good agreement with the migration time
appropriate to a fully three dimensional locally isothermal disc, and that two 
dimensional simulations can be used to study the migration of low mass 
protoplanets provided  an appropriate 
gravitational softening  parameter is used.

Figure~\ref{fig2} shows the migration time calculated for simulations
in which $\Sigma(r) \propto r^{-1}$, $H/r=0.07$, and the softening
parameter was $b=0.3 H_p$. These are the values adopted for the
global turbulent disc runs presented in sections~\ref{G1}, \ref{G2}, and 
\ref{G3}, because the potential is better represented close to the
planet. The disc model here was normalised so that it contained
the equivalent of $0.02$ M$_{\odot}$ between 0 and 40 AU, and $\simeq 2.5$
Jupiter masses interior to the protoplanet radius $r_p$. Figure~\ref{fig2}
shows that  this results
in migration times that are a factor of $\simeq 2/3$ shorter than those
predicted by Tanaka, Tacheuchi, \& Ward (2002) for the smaller
masses but are in slightly closer agreement in the higher $\sim 30 M_{\oplus}$
range. These shorter migration times arise because of the smaller softening
parameter employed.

\subsection{Global Model G1} \label{G1}
\begin{figure}
\centerline{
\epsfig{file=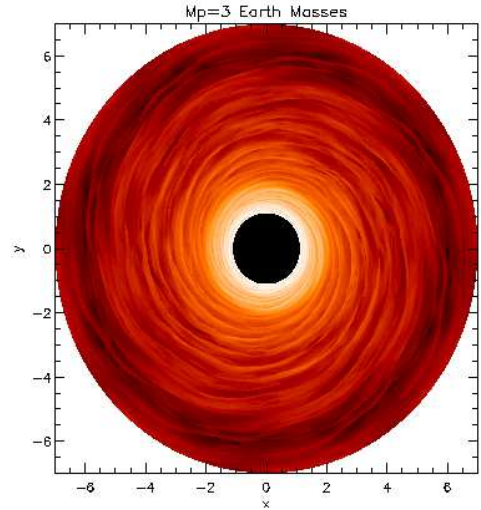,width=\columnwidth} }
\caption[]
{This figure shows midplane density contours for the run G1. Note that the
presence of the protoplanet is undetectable due to the higher amplitude
perturbations generated by the turbulence.
The protoplanet is located at ($x_p$,$y_p$)=
(-3,0).}
\label{fig4}
\end{figure}
\begin{figure*}
\centerline{
\epsfig{file=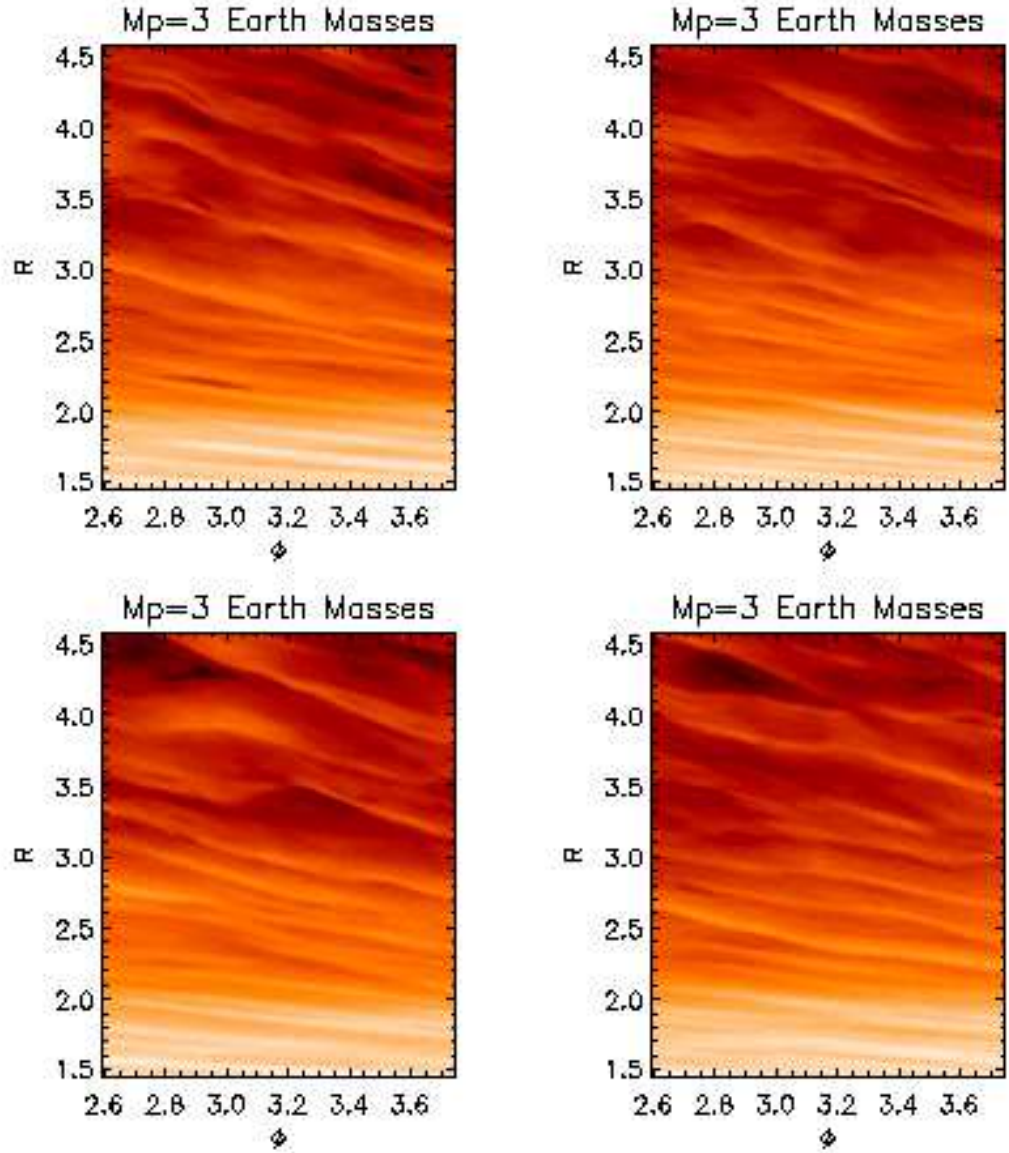,width=\textwidth} }
\caption[]
{This figure shows midplane density contours for the run G1 in the region
close to the protoplanet. The protoplanet is located at ($r_p$,$\phi_p$)=
(3,$\pi$). The panels correspond to times 452.4903, 497.7334, 542.9972, and
672.4058, respectively. Note that the
presence of the protoplanet is undetectable due to the higher amplitude
perturbations generated by the turbulence.}
\label{fig5}
\end{figure*}
\begin{figure}
\centerline{
\epsfig{file=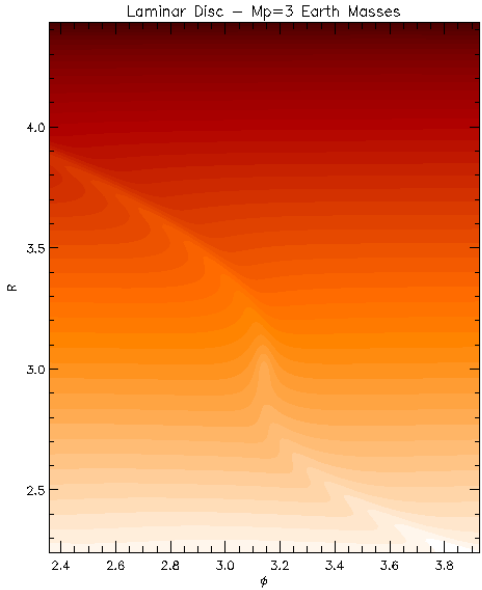,width=\columnwidth} }
\caption[]
{This figure shows midplane density contours for a laminar disc run
with planet parameters identical to run G1.
The presence of the planet due to the excitation of spiral waves
is clearly detectable, in contrast to the picture presented in figure~\ref{fig5}.}
\label{fig6}
\end{figure}

The global model G1 described in table~\ref{table1} contained a protoplanet
whose mass is equivalent to $\simeq 3$ Earth masses. 
Such a protoplanet is expected
to provide a small perturbation to the disc. A snapshot of the midplane
density is shown in figure~\ref{fig4}, and shows that the presence of
the protoplanet is indiscernible due to the higher amplitude density 
fluctuations generated by the turbulence. Figure~\ref{fig5} shows a
series of close--up images of the midplane density in the near vicinity of
the protoplanet, which again show that the protoplanet cannot be detected.
A similar plot for an equivalent laminar disc run is presented in 
figure~\ref{fig6}
for purposes of comparison. It is worth noting that once the density structure
shown in figure~\ref{fig6} is established in the laminar disc runs,
it remains essentially time independent. The changing density structure observed
in figure~\ref{fig5}, combined with the larger amplitude of the
turbulent wakes when compared with those generated by the protoplanet, suggest
that the gravitational interaction between a low mass protoplanet and
turbulent disc  may differ substantially when compared with a laminar disc.

\begin{figure}
\centerline{
\epsfig{file=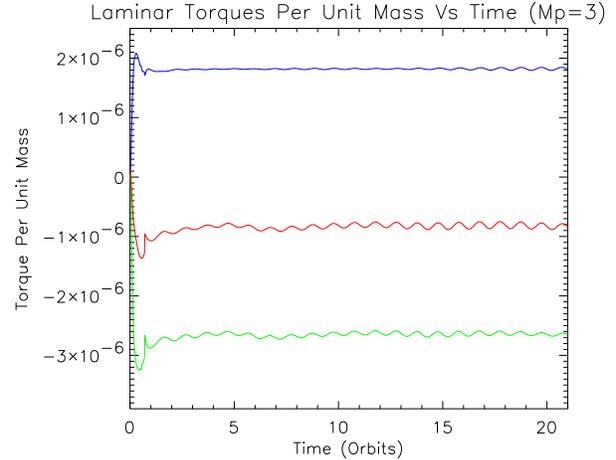,width=\columnwidth} }
\caption[]
{This figure shows the torque per unit mass exerted by the disc
on the protoplanet for a laminar disc and planet parameters
identical to run G1. The upper line shows the torque due to the
inner disc, the lower line shows the torque due to the outer disc,
and the middle line shows the total torque. It is clear that a well defined
net torque is produced, with an associated migration time scale.}
\label{fig7}
\end{figure}

Figure~\ref{fig7} shows the time evolution of the
torque per unit mass exerted on a protoplanet
of equivalent parameters to that in run G1 but embedded in a laminar disc.
The upper line shows the (positive) torque due to the disc lying interior
to the protoplanet orbital radius, the lowest line shows the torque
due to the outer disc, and the middle line shows the total (negative) torque.
It is clear that a well defined net torque acting on the protoplanet may be
defined, along with a corresponding inward migration rate.

\begin{figure}
\centerline{
\epsfig{file=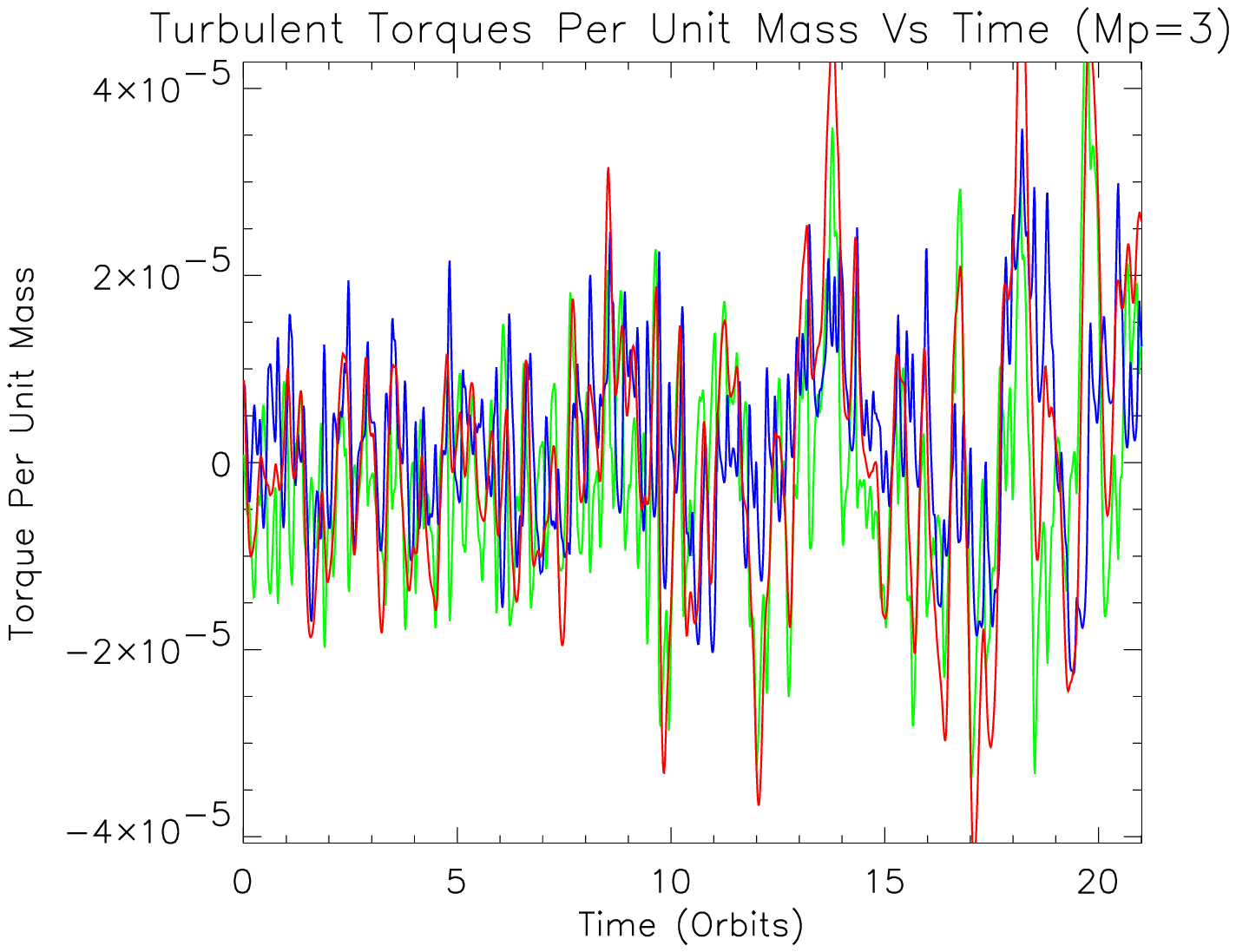,width=\columnwidth} }
\caption[]
{This figure shows the torque per unit mass exerted by the disc
on the protoplanet for the run G1. The turbulence in this case
generates strong fluctuations in the torque from each side of the disc, 
such that it becomes difficult to distinguish the torques arising from
each side. The total torque fluctuates between positive and negative 
values such that the associated migration will undergo a `random walk'.}
\label{fig8}
\end{figure}

Figure~\ref{fig8} shows a similar plot but for the turbulent disc model
G1. It is clear that the forcing experienced by the protoplanet in this
case differs dramatically from that obtained in the
laminar disc. 
The torques from the inner and outer disc suffer very large fluctuations as
a result of the protoplanet interacting gravitationally with the turbulent wakes
apparent in figures~\ref{fig4} and \ref{fig5} as they shear past the
protoplanet, and this causes the net torque
experienced by the protoplanet to oscillate between negative and positive
values. The orbital migration of the protoplanet is thus likely to 
occur as a random walk rather than as a monotonic inward drift normally 
associated with type I migration.
In this run, for which the spiral wakes generated by the planet
are completely dominated by the turbulent wakes, the usual separation
between inner and outer disc torques is barely discernible 
due to the high amplitude turbulent fluctuations.

\begin{figure}
\centerline{
\epsfig{file=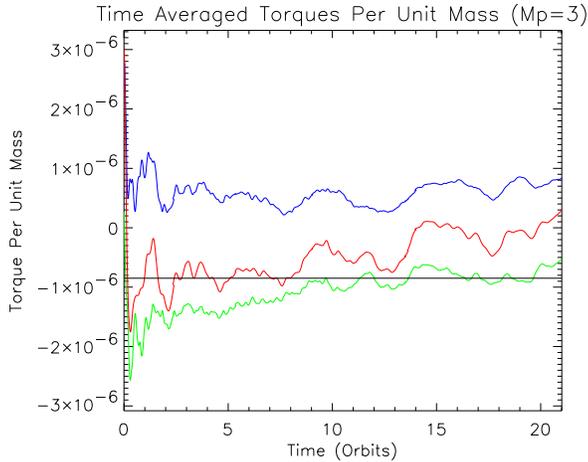,width=\columnwidth} }
\caption[]
{This figure shows the running time average of the
torque per unit mass exerted by the disc for run G1. The upper line is the 
running time average of the torque acting on the planet due to the inner disc.
The lower line is that due to the outer disc. The middle (not straight) line
is the running time average of the total torque. The straight line is the
total torque exerted on the protoplanet in a laminar disc run. We note that
the total time averaged torque does not converge to a well defined value.}
\label{fig9}
\end{figure}

The time evolution of the running time average of the torque per unit mass
in model G1 is shown in figure~\ref{fig9}. The upper line shows
the running time average of the torque due to the inner disc, the
lowest line shows the time averaged torque due to the outer disc, and
the middle line shows the running time average of the total torque.
The straight line shows the time average of the total torque per unit mass
experienced in an equivalent laminar disc
model. The first thing to note from figure~\ref{fig9} is that the running time
averaged torque in run G1 does not converge to a well defined value
for the time over which the
simulation was run (just over 20 planet orbits). 
Near the beginning of the simulation the time average is close to that obtained from a similar laminar
disc run, but as the calculation evolves the running time average 
changes continuously. By the end of the simulation,
the torque experienced by the protoplanet
will be positive on average, corresponding to outward migration on a time scale
of $\simeq 2.8$ Myr.

If we assume that there exists a well defined long term mean for the
torque experienced by the protoplanet, then we can express the torque
as the sum of this mean torque and a fluctuating component:
\begin{equation}
T(t) = {\overline T} + T_f(t) \label{torque-mean1} 
\end{equation}
where $T(t)$ is the torque experienced at time $t$, ${\overline T}$ is 
the mean torque, and $T_f(t)$ is the fluctuating component.
The running time average of the torque is given  by
\begin{equation}
\langle T \rangle_t = \frac{\int_0^t T(t) dt}{\int_0^t dt} \label{T_av}
\end{equation}
then we can write
\begin{equation}
\langle T \rangle_t = {\overline T} + \frac{1}{t} \int_0^t T_f(t) dt. 
\label{T_av2}
\end{equation}
Assuming that the amplitude of the torque fluctuations 
$T_f(t)$ about the mean $\overline T$
follow a Gaussian distribution with standard deviation $\sigma_T$
and recur on a characteristic timescale $t_f,$
we can estimate from 
equation~\ref{T_av2}  that the fluctuations in the running
mean satisfy
\begin{equation}
| \langle T_t  \rangle - {\overline T}| \sim  \sigma_T\sqrt{{t_f \over t}}, 
\label{T_av3}
\end{equation}
which may be used to estimate how long we need to run a simulation before
we can expect the running time average of the torque to converge
to a well defined value (i.e. when the  fluctuation estimate from
equation~\ref{T_av3}  becomes significantly less than $|{\overline T}|$).
 Inspection of
figure~\ref{fig8} suggests that the  $t_f$ is
 is typically about one half of a planetary orbit, $P_p/2$.
Equation~\ref{T_av3} tells us that
the larger the relative 
amplitude of fluctuations, the longer we have to integrate
for the running mean to converge.

It is clear that the level of fluctuations associated with the torques
is very much larger than the running mean in figure~\ref{fig8}.
If we take the mean value of the torque to be 
${\overline T} \simeq 9 \times 10^{-7}$ from the straight 
line in fig~\ref{fig8},
then the typical magnitude of the fluctuations can be estimated to be
$\sigma_T \simeq 2 \times 10^{-5}$, but with extreme fluctuations being a factor
of $\sim 40$ times larger than the mean
during certain periods of the evolution. If we set $\sigma_T \simeq 2 \times
10^{-5}$, then the expected time for convergence is approximately
250 planetary orbits. 
This time scale is much longer than 
the time over which it is feasible to run simulations of this kind at the
present time, and may provide at least a partial explanation of
why convergence is not obtained in figure~\ref{fig9}.
But note that it may give a significant underestimate for the time
required for convergence if the statistics of the fluctuations are
more complex involving say a superposition of fluctuations
with varying characteristic times up to the total evolution
time of the global disc. In such a situation convergence to small
means may be practically impossible.

\subsection{Global Model G2} \label{G2}

\begin{figure}
\centerline{
\epsfig{file=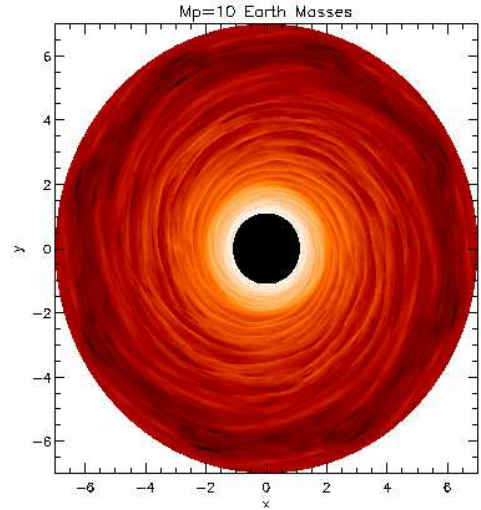,width=\columnwidth} }
\caption[]
{This figure shows midplane density contours for the run G2. Note that the
presence of the protoplanet is just detectable although the 
perturbations generated by the turbulence are of higher amplitude than those
generated by the protoplanet.
The protoplanet is located at ($x_p$,$y_p$)=
(-3,0).}
\label{fig10}
\end{figure}

\begin{figure*}
\centerline{
\epsfig{file=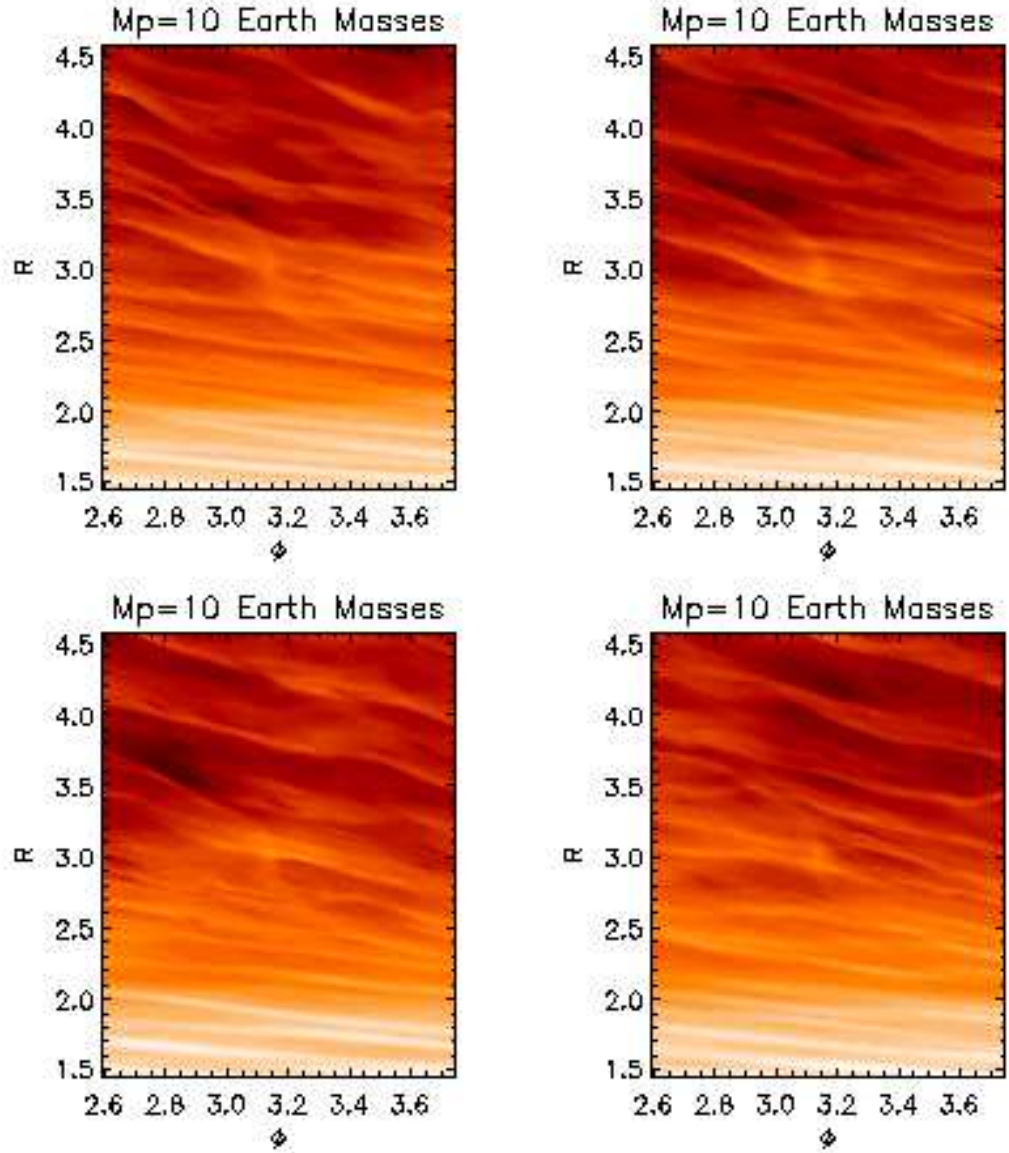,width=\textwidth} }
\caption[]
{This figure shows midplane density contours for the run G2 in the region
close to the protoplanet. The protoplanet is located at ($r_p$,$\phi_p$)=
(3,$\pi$). The panels correspond to times 429.0687, 496.8698, 542.0710, and
634.4136, respectively. Note that the
presence of the protoplanet is just detectable, although the perturbation
it makes to the disc is of lower amplitude than the 
perturbations generated by the turbulence.}
\label{fig11}
\end{figure*}

\begin{figure}
\centerline{
\epsfig{file=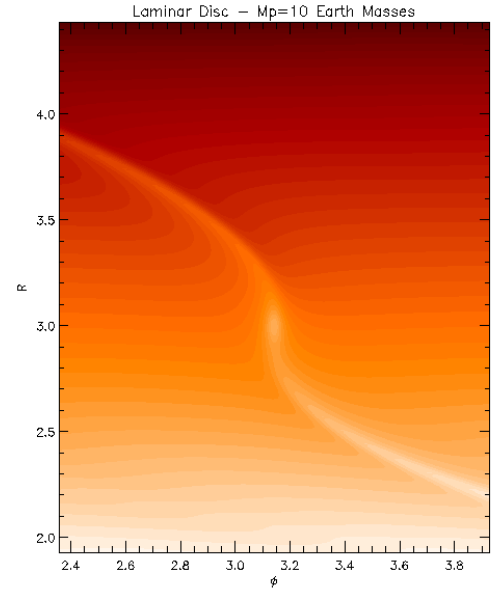,width=\columnwidth} }
\caption[]
{This figure shows midplane density contours for a laminar disc run 
with planet parameters identical to run G2. This figure should be
compared with figure~\ref{fig11}.}
\label{fig12}
\end{figure}

The protoplanet mass in this case was equivalent to $\simeq 10$ Earth masses.
A snap shot of the midplane density distribution generated by run G2 is shown
in figure~\ref{fig10}. The presence of the protoplanet is just
discernible at ($x$, $y$)= (-3.,0). Close up images of the density
field in the vicinity of the protoplanet are shown in figure~\ref{fig11},
and again the presence of the protoplanet can just be detected along with the
spiral waves excited by the protoplanet. The time variability of the density
field close to the protoplanet is apparent in this figure.
Figure~\ref{fig12} shows the density field in the vicinity of the
protoplanet in a laminar disc run that is otherwise equivalent to run G2,
and it is worth noting that the density field shown in figure~\ref{fig12}
remains essentially
time independent once established.

\begin{figure}
\centerline{
\epsfig{file=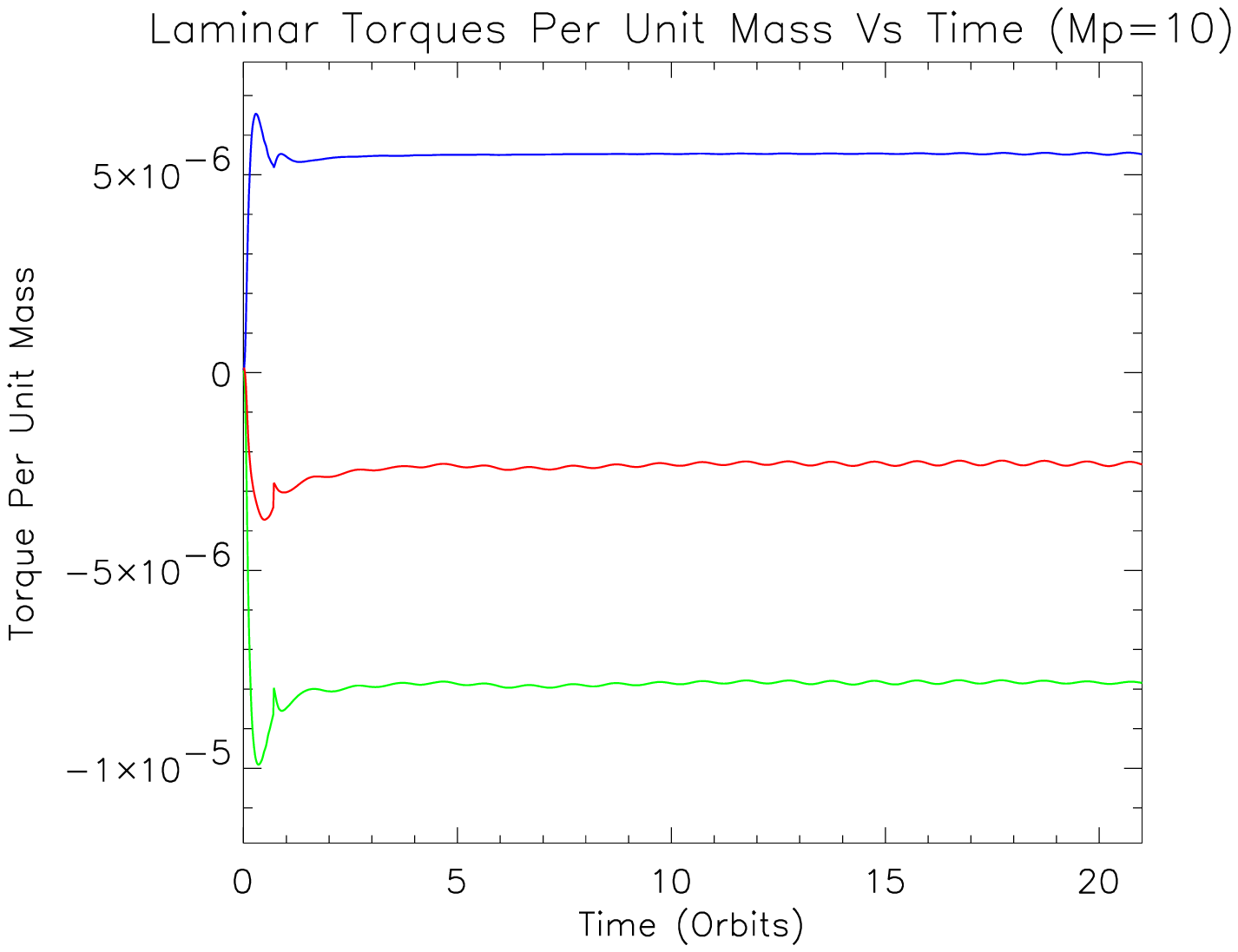,width=\columnwidth} }
\caption[]
{This figure shows the torque per unit mass exerted by the disc
on the protoplanet for a laminar disc and planet parameters
identical to run G2. The upper line shows the torque due to the
inner disc, the lower line shows the torque due to the outer disc,
and the middle line shows the total torque. It is clear that a well defined
torque is produced, with an associated migration time scale.}
\label{fig13}
\end{figure}

\begin{figure}
\centerline{
\epsfig{file=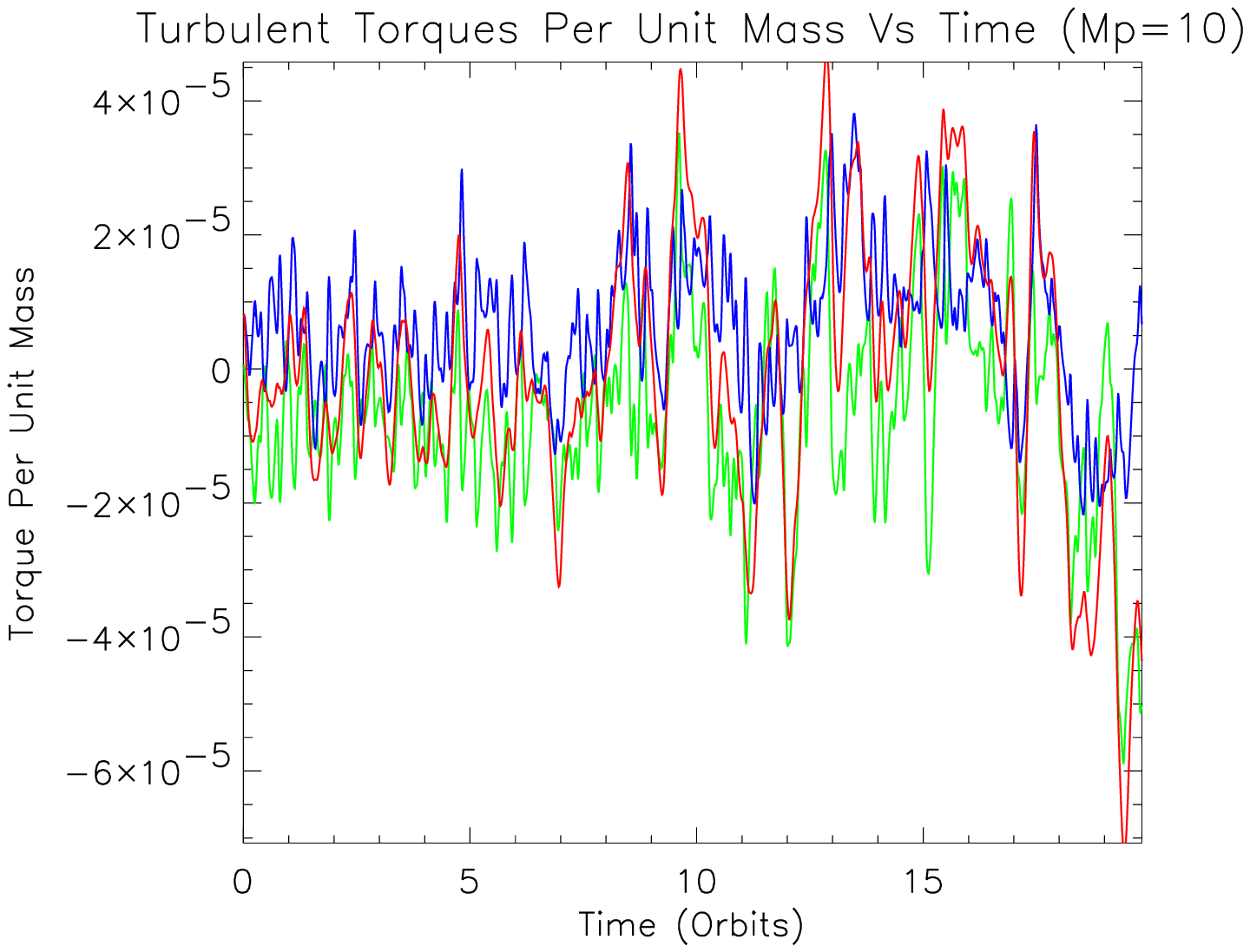,width=\columnwidth} }
\caption[]
{This figure shows the torque per unit mass exerted by the disc
on the protoplanet for the run G2. The turbulence in this case
generates strong fluctuations in the torque from each side of the disc, 
such that it becomes difficult to distinguish the torques arising from
each side. The total torque fluctuates between positive and negative 
values such that the associated migration will undergo a `random walk'.}
\label{fig14}
\end{figure}

Figure~\ref{fig13} shows the torque per unit mass as a function of
time for a laminar run equivalent to G2. The approximate
constancy of the middle
line in this plot (which is the total torque per unit mass on the
protoplanet) shows that a well defined torque and migration time 
can be ascribed in this case.  Figure~\ref{fig14} shows an equivalent plot
for run G2. As in run G1, the torque evolution in this case is
characterised by large fluctuations about a small mean, such that
the orbital evolution of the protoplanet is likely to occur as a random walk.

\begin{figure}
\centerline{
\epsfig{file=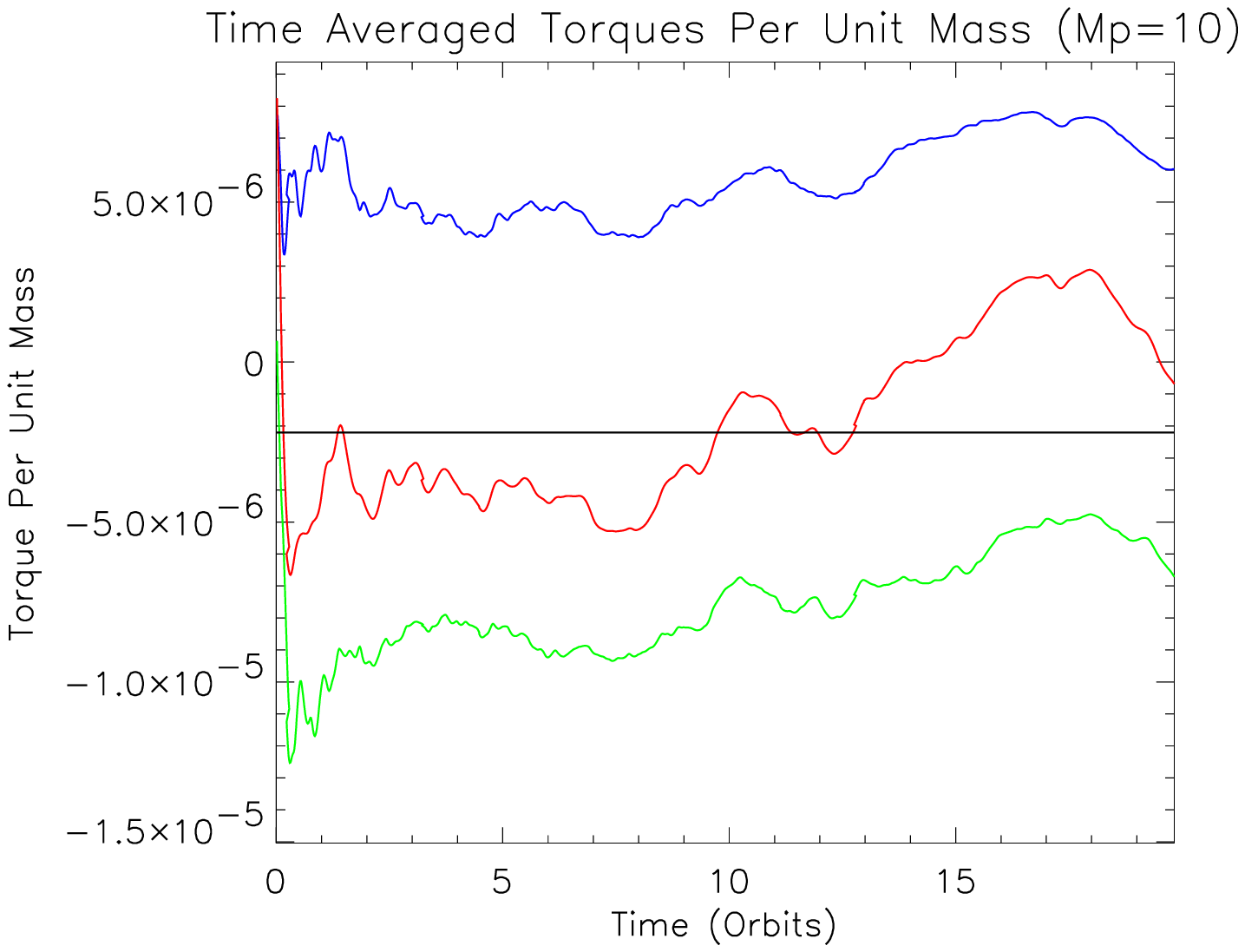,width=\columnwidth} }
\caption[]
{This figure shows the running time average of the
torque per unit mass exerted by the disc for run G2. The upper line is the 
running time average of the torque acting on the planet due to the inner disc.
The lower line is that due to the outer disc. The middle (not straight) line
is the running time average of the total torque. The straight line is the
total torque exerted on the protoplanet in a laminar disc run. We note that
the total time averaged torque does not converge to a well defined value.}
\label{fig15}
\end{figure}

The running time average of the torques for run G2 are plotted in 
figure~\ref{fig15}. The time average of the total torque is again
found to not converge, showing variations that indicate inward migration on
average after $\simeq 7$ orbits have elapsed, outward migration on average
after $\simeq 18$ orbits, and inward migration on average 
again at the very end of the simulation. Using similar arguments presented
in section~\ref{G1} to estimate the time for the running mean to converge,
we again estimate that the run time required will be around 70 -- 80 planet
orbits. This is significantly more than the run time over which we are
able to compute these models at the present time.  However,
the torques from the two sides of the disc exhibit somewhat relatively
smaller amplitude fluctuations when compared to simulation G1.
This is reflected in the shorter time estimated for convergence.
It is also part of a tendency to have smaller relative fluctuations
for larger protoplanet masses which makes net torque measurement
easier. The same trend is seen in the shearing box simulations (see below).
We also comment that the typical magnitude of the average torques from
each side of the disc corresponds to the laminar value even in the 
low mass cases
in spite of the large noise levels.

Although  that may be the case,
large noise levels which apparently can occur on
a variety of time scales will cause the migration
 of a low mass protoplanet embedded
in a turbulent disc  to differ substantially from that found 
in a laminar disc model,
with periods of inward migration interspersed with outward migration.
At this present time we are unable to give a reliable estimate of
the net migration time of protoplanets in such disc models, or 
because there may be significant variations occurring on the 
longest evolutionary time scales of the global disc, give 
even a
reliable indication of the direction of migration in the long term.
Feedback of the orbital evolution on the turbulence may
be significant in some cases.
Numerical experiments are currently being performed to examine this
 and will be the subject of a future publication.

In a recent paper, Winters, Balbus, \& Hawley (2003a)
have emphasised the `chaotic' nature of MHD turbulence, with the implication
being that the evolution of the turbulent state may be modified by small
perturbations. The presence of a protoplanet induces such perturbations.
Both simulations G1 and G2 were initiated with identical disc models (but not
run G3),
as commented upon in paper III. In paper III we noted that the global
magnetic energy of the system was found to diverge (by a modest amount)
once the planets had been inserted in the disc models. By comparing figure
\ref{fig8} with \ref{fig14}, and \ref{fig9} with \ref{fig15}, we can see the
affect that the differing planet masses have on the local statistics
of the turbulence. While the overall magnitude and form of the
torque fluctuations are similar, the details are clearly different.
Given that these fluctuations appear to strongly influence the
orbital evolution of the protoplanets, we can conclude that the detailed
orbital evolution of a low mass  protoplanet in  a turbulent disc
is essentially indeterminate
due to the feedback between the turbulence and the planetary perturbations.
Interestingly enough though, we can say that even if the mean average
torque results in inward migration within a corresponding expected  lifetime
in the disc,  large random  fluctuations will
result in some  increased survival probability for such a time  for a subset
of an ensemble of protoplanets.

\subsection{Global Model G3} \label{G3}
\begin{figure}
\centerline{
\epsfig{file=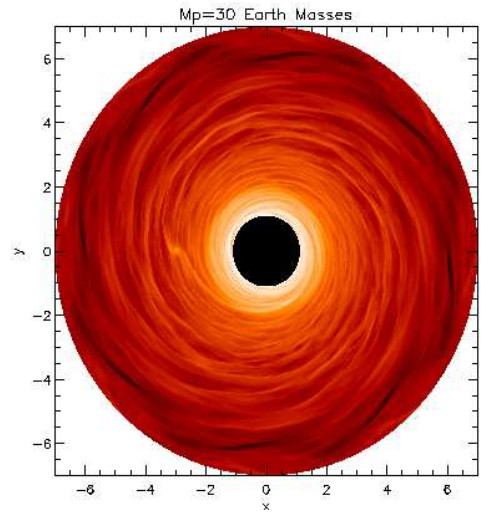,width=\columnwidth} }
\caption[]
{This figure shows midplane density contours for the run G3. The
presence of the protoplanet is  clearly detectable, with the perturbations
generated by the protoplanet being of similar magnitude to those
generated by the turbulence.
The protoplanet is located at ($x_p$,$y_p$)=
(-3,0).}
\label{fig16}
\end{figure}

\begin{figure*}
\centerline{
\epsfig{file=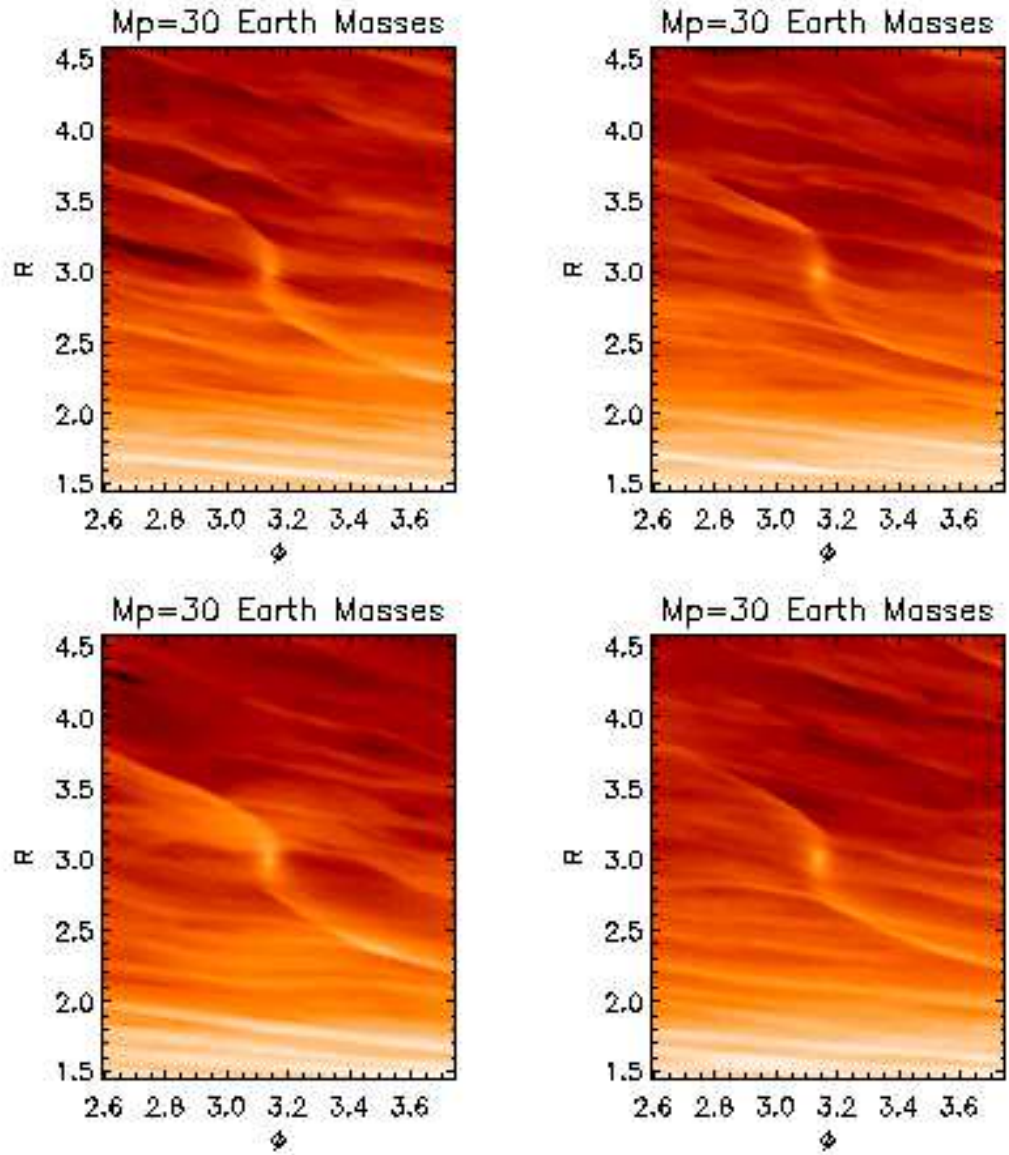,width=\textwidth} }
\caption[]
{This figure shows midplane density contours for the run G3 in the region
close to the protoplanet. The protoplanet is located at ($r_p$,$\phi_p$)=
(3,$\pi$). The panels correspond to times 871.4805, 903.0731, 948.2260, and
1077.837, respectively. Note that the
presence of the protoplanet is clearly detectable, with the perturbation
it makes to the disc being of similar amplitude to the 
perturbations generated by the turbulence. Note that the appearance of
the spiral waves generated by the protoplanet are time variable since they
are perturbed by the turbulence.}
\label{fig17}
\end{figure*}

\begin{figure}
\centerline{
\epsfig{file=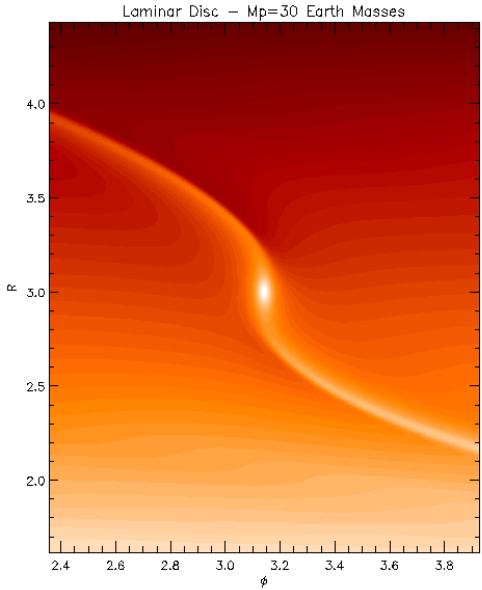,width=\columnwidth} }
\caption[]
{This figure shows midplane density contours for a laminar disc run
with planet parameters identical to run G3. This figure should be
compared with figure~\ref{fig17}.}
\label{fig18}
\end{figure}

Figure~\ref{fig16} shows a snap shot of the midplane
density structure in model G3,
for which the protoplanet mass is equivalent to 30 Earth masses.
The protoplanet is located at ($x$,$y$)=(-3.,0), and is clearly visible,
along with the spiral waves it has generated. Figure~\ref{fig17}
shows a series of close--up images of the midplane density near the 
protoplanet. Although the protoplanet and spiral waves it generates are
clearly visible, it is also apparent that the density structure near the
planet is highly variable. A close--up image of the density near the protoplanet
for an equivalent laminar disc run is shown in figure~\ref{fig18}
for comparison.
The density field and spiral wakes in this case are essentially time 
independent.

\begin{figure}
\centerline{
\epsfig{file=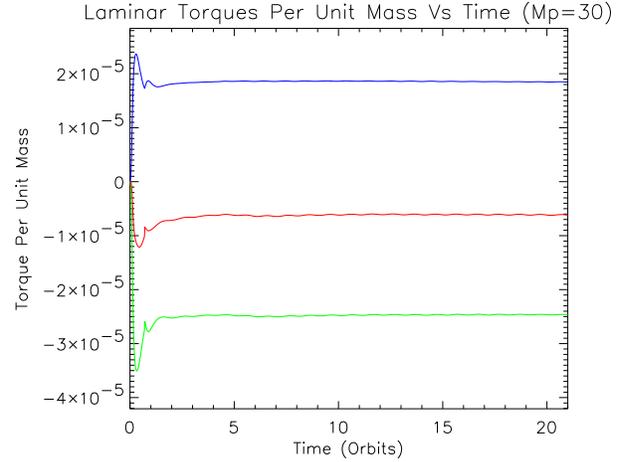,width=\columnwidth} }
\caption[]
{This figure shows the torque per unit mass exerted by the disc
on the protoplanet for a laminar disc and planet parameters
identical to run G3. The upper line shows the torque due to the
inner disc, the lower line shows the torque due to the outer disc,
and the middle line shows the total torque. It is clear that a well defined
torque is produced, with an associated migration time scale.}
\label{fig19}
\end{figure}

\begin{figure}
\centerline{
\epsfig{file=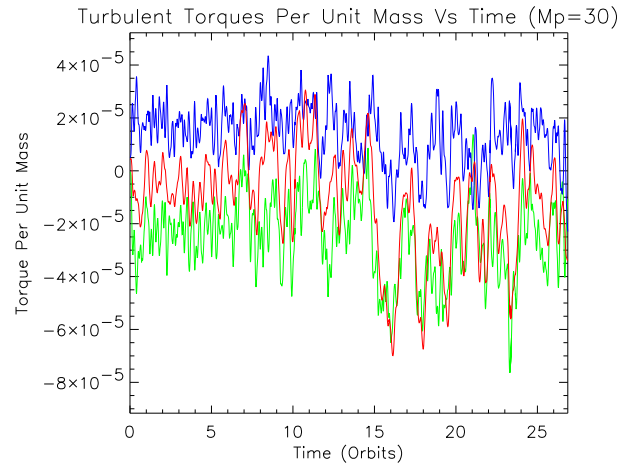,width=\columnwidth} }
\caption[]
{This figure shows the torque per unit mass exerted by the disc
on the protoplanet for the run G3. The turbulence in this case
generates strong fluctuations in the torque from each side of the disc, 
such that it becomes difficult to distinguish the torques arising from
each side. The total torque fluctuates between positive and negative 
values such that the associated migration will undergo a `random walk'.}
\label{fig20}
\end{figure}

The torque per unit mass as a function of time is shown in figure~\ref{fig19}
for the laminar disc run equivalent to run G3. The middle line shows the
net torque per unit mass, whose approximate
constancy leads to a well defined 
migration time. The time evolution of the torque per unit mass
for run G3 is shown in figure~\ref{fig20}. A similar situation to that
described for runs G1 and G2 is observed, with the torques being 
significantly modified by strong fluctuations. Again the total torque
can be seen to show periods of being positive and negative, suggesting that
the protoplanet would undergo a random walk rather than monotonic,
inward migration. 

We note, however, that as the protoplanet mass increases, and the spiral
waves excited increase in amplitude to become comparable or larger that
the turbulent density wakes, the torques due to the inner and outer
disc begin to separate. This can be observed by comparing figures~\ref{fig8},
\ref{fig14}, and \ref{fig20}, which show the torques
from the inner and outer discs becoming progressively distinguishable from
each other as the
planet mass increases. This is also accompanied by a reduction
in the relative  torque fluctuation noise level.

\begin{figure}
\centerline{
\epsfig{file=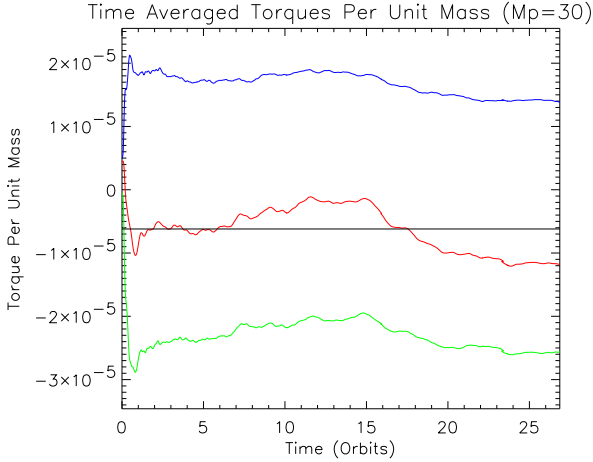,width=\columnwidth} }
\caption[]
{This figure shows the running time average of the
torque per unit mass exerted by the disc for run G3. The upper line is the 
running time average of the torque acting on the planet due to the inner disc.
The lower line is that due to the outer disc. The middle (not straight) line
is the running time average of the total torque. The straight line is the
total torque exerted on the protoplanet in a laminar disc run. We note that
the total time averaged torque does not converge to a well defined value.}
\label{fig21}
\end{figure}

Figure~\ref{fig21} shows the running time average of the torques
per unit mass for run G3. The straight line in this figure is the time averaged
total
torque per unit mass
obtained in the equivalent laminar disc run. As in runs G1 and G2, the
running time average of the total torque per unit mass fails to
converge for the run time considered here.
However, the relative fluctuation amplitudes are smaller in this case
leading to a smaller anticipated time for convergence.
This is part of the  trend for larger mass protoplanets to produce
larger amplitude perturbations that are more difficult for the turbulence to
affect.

If we consider the running mean of the total torque 
in figure~\ref{fig21} we can reasonably take a value of
${\overline T} \simeq - 6 \times 10^{-6}$. A by--eye inspection of 
figure~\ref{fig20} indicates that the level of fluctuation in the torques
is $\sigma_T \simeq 4 \times 10^{-5}$. The predicted run time
for convergence of the running mean from equation~\ref{T_av3} is then
$\simeq 20$ -- $30$ planetary orbits, 
which is shorter than for G1 or G2 and is comparable
to  the time for which the simulation
has been run. 

Although it always corresponds 
to inward migration, the non convergence of the running mean toward a well
defined value suggests that a simple picture of local turbulence
in the vicinity of the planet, in which the state variables have well
defined mean values on top of which are superposed fluctuations
with a well defined Gaussian spectrum, is not accurate. Instead, it appears that
the global nature of the disc plays an important role in continuously
modifying the local structure of the disc and turbulence. Communication
between different regions of the disc can affect the local properties
such as the mean density, the amplitude, spatial, and temporal
distribution of density and velocity fluctuations over long time scales,
such that a local
running mean
is problematic to define. An examination of the global properties of
a turbulent disc model, such as the global magnetic energy (see e.g. 
papers I--III) show that there are short
and longer time scale variations in the turbulence that reflect modifications
to the local and global turbulence, and by implication the local
disc structure. The results presented in section~\ref{box-res}
for the local shearing box simulations show smaller relative fluctuations
and greater convergence of the mean
torques toward
the expected value, indicating that the global properties of turbulent
discs may play an important role by inducing longer time scale
modification to the local state of the disc in the vicinity of an embedded
protoplanet.

\subsection{Global Run G5}
The planet mass in global run G5 was equivalent to 3 Jupiter masses.
The computational domain in this case was restricted so that the azimuthal
interval ran between $[0,\pi/2]$ (thus allowing a reasonable run time),
with the protoplanet being placed
on a fixed circular orbit at $(r_p,\phi_p)=(2.5,\pi/4)$.
The physical parameters were identical to run G4, for which the azimuthal
domain covered the full $2 \pi$. However, run G4 was only run for $\simeq 11$
planetary orbits, which is too short a time for anything other than the
tendency for gap formation and the global disc morpholgy to be inferred.
Consequently run G4 will not be discussed further here.

\begin{figure}
\centerline{
\epsfig{file=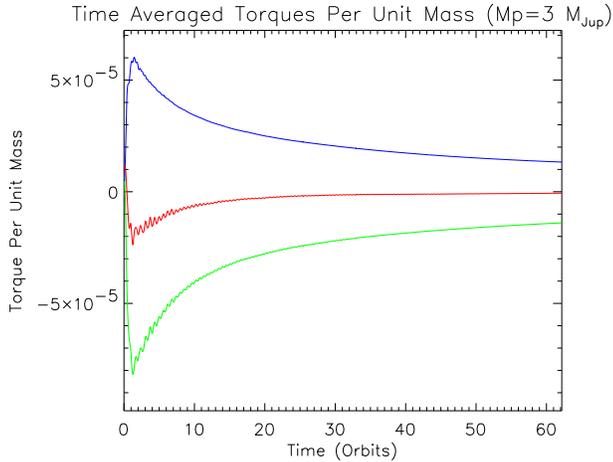,width=\columnwidth} }
\caption[]
{This figure shows the running time average of the
torque per unit mass exerted by the disc for run G5. The upper line is the
running time average of the torque acting on the planet due to the inner disc.
The lower line is that due to the outer disc. The middle line
is the running time average of the total torque.
The running time average for both the outer and inner torque converge to well
defined values in this run, for which the interaction is non linear and
the planet opens up a large gap. We note that our use of a closed inner 
boundary 
causes the density of the inner disc to be artificially increased after 
gap formation, such that the torque 
due to the inner disc is probably an over estimate,
leading to a larger torque contribution.}
\label{fig22}
\end{figure}

Run G5 has been
described in paper III, and showed a tendency toward clear gap formation
with the response of the disc due to the presence of the planet being strongly
non linear. Consequently the perturbations induced in the disc by the
protoplanet
are very much larger than those that arise because of the
turbulence. This results in the fluctuations in the torque experienced by the
protoplanet being significantly smaller (in relative terms)
than observed in the previously
described runs G1, G2, and G3,
and a well defined running time average of the torque
being obtained. Figure~\ref{fig22} shows the running time average of
the torque per unit mass obtained from run G5, with the upper line
corresponding to the 
inner disc torque, the lowest line corresponding to the outer
disc torque, and the middle line the running time average of the total torque.
It is clear from this figure that a large torque is exerted on the protoplanet
prior to gap formation, but that as the gap proceeds to open and material
is pushed away from the planet the torque
diminishes. The running time averaged
torques due to the inner and outer disc appear to be approaching well defined
asymptotic values, which are unaffected by turbulent fluctuations,
but the continued decrease in the running time averages indicates
that gap formation is still ongoing at the end of the simulation.

We note that the use of a closed inner boundary in this simulation, combined
with the close proximity of the planet to the inner boundary, cause the
density of the inner disc to be maintained at an artificially high level
after gap formation. This leads to the net torque on the planet being
negative but
close to zero. Under more general circumstances in which the inner disc
can accrete onto the central star, an inner cavity is expected to form
such that the torque on the protoplanet is dominated by the outer
disc (e.g. Nelson et al. 2000). If we adopt the disc model described
in section~\ref{calibration} used to normalised the results 
presented in figure~\ref{fig2}, 
and estimate the migration time using equation~\ref{tmig-sim} and a torque 
per unit mass
due to the outer disc in figure~\ref{fig22} of $T=-10^{-5}$, 
then we obtain
$\tau_{mig} \simeq 4 \times 10^4$ yr, for a planet at 5.2 AU. The type II
migration time appropriate to gap forming protoplanets is given
by the viscous evolution time
$\tau_{mig}= (2r_p^2)/(3 \nu)$ where $\nu=\alpha H^2 \Omega$ is the
kinematic viscosity. For a disc model with $\alpha \simeq 7 \times 10^{-3}$
and $H/r=0.07$, the estimated type II migration time is
$\tau_{mig}=4.5 \times 10^4$ yr, in reasonable agreement with the result
obtained from the simulation G5. 

We note that the above estimates for type II migration times
correspond to disc models with full $2 \pi$ azimuthal domains, and that it is
unclear which precise value the running mean of the outer disc torque will
approach once gap formation is complete. Nonetheless, the reasonable agreement
obtained in the estimates suggests that gap forming protoplanets in turbulent 
discs undergo migration at the expected type II rate.
A similar result was obtained 
in paper II for type II migration rates in turbulent discs with
full $2 \pi$ azimuthal domains.

It is clear that a well defined trend arises when considering the interaction
between embedded protoplanets and turbulent discs. Lower mass objects that
are unable to perturb the turbulent background flow significantly
are subject to strong torque fluctuations that are likely to dominate
their orbital evolution. As the protoplanet mass increases so that
the amplitude of the spiral wakes that it excites become
larger than the turbulent density fluctuations, the relative
magnitudes of the torque fluctuations decrease, and the migration is 
likely to
become similar to type I migration (although with a significant noise
component). For larger protoplanet masses that allow gap formation, the
effect of the turbulent fluctuations is small, with the migration
being essentially the same as the standard type II picture.
These trends are also observed in the shearing box simulations that are
described below.

\subsection{Shearing Box Simulation Results} \label{box-res}
\begin{figure*}
\centerline{
\epsfig{file= 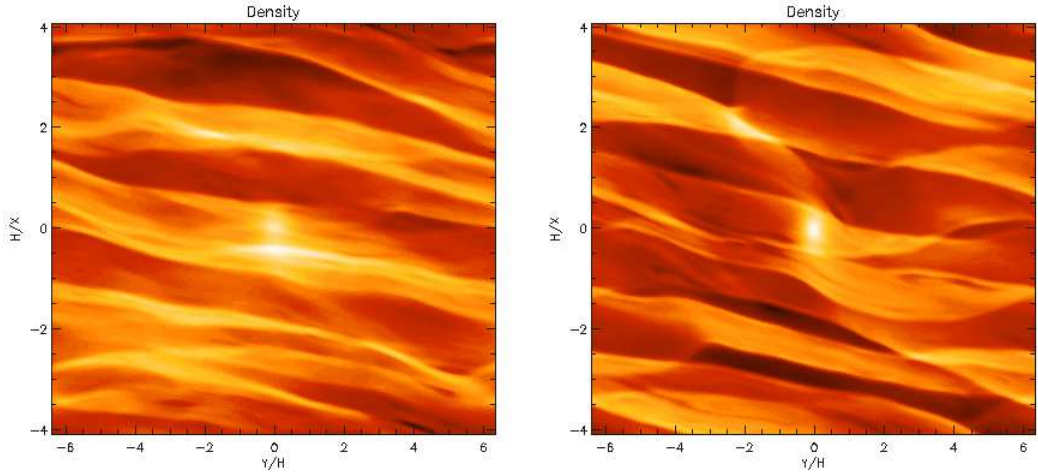,width=\textwidth} }
\caption[]
{Density contour plots taken in the box mid plane
for simulations Ba1 and Ba2 which have the smaller mass and hence
fully embedded protoplanets. The wakes  produced by these protoplanets
can be discerned even though the medium is turbulent.
}
\label{figbox1}
\end{figure*}

\begin{figure*}
\centerline{
\epsfig{file=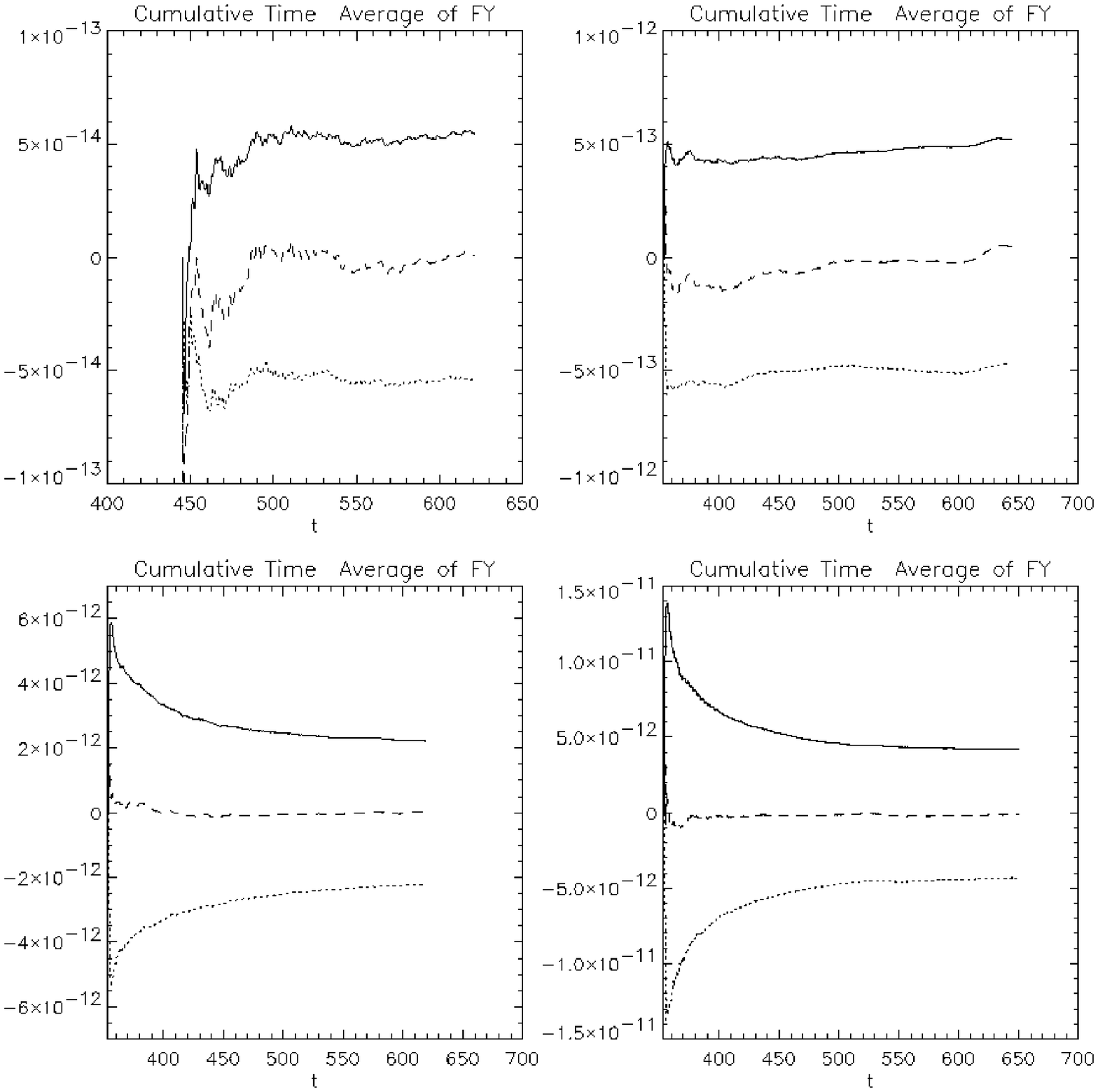,width=\textwidth} }
\caption[]
{This figure shows running time averages of the torques acting on the
protoplanet in simulations  Ba1 top left panel, Ba2 top right panel,
Ba3 bottom left panel and Ba4 bottom right panel. In each panel the
upper lower and middle curves correspond to running averages
of the torques acting on the planet due to
the regions of the box interior to the planet, the torque due to
the regions exterior to the planet, and the net torque respectively.
}
\label{figbox2}
\end{figure*}

Details of the 
shearing box simulations Ba1 - Ba4 are given in 
table~\ref{table2}. These  were each
continued from a simulation with fully developed
turbulence Ba0 after inserting a protoplanet
with values of the dimensionless parameter
$GM_p /( H^3\Omega_p^2) = M_p R^3/(M_* H^3)$ measuring
the mass of the protoplanet  equal
to $0.1, 0.3, 1$ and $2$ respectively. In this
section, $\Omega_p$ and $R$ are the angular velocity
and radius of the centre of the box. Thus simulations
Ba1 and Ba2 are directly comparable to the global simulations
G2 and G3 in terms of the relative strength of the protoplanet's
perturbation on the disc. As in those cases the protoplanets
remain embedded with no indication of gap formation.
However, simulations Ba3 and Ba4 show gap formation and a non linear
perturbation of the disc as do simulations G4 and G5.
We here discuss the force exerted on the protoplanet
by the disc. In particular we study the component in the azimuthal
direction $F_y.$ Adopting a unit of length equal to
the radius of the center of the box, this is also the torque.

Density contour plots taken in the box mid plane
  near the end of the 
simulations Ba1 and Ba2 which have 
embedded protoplanets are shown in figure \ref{figbox1}. 
As in the global simulations, the wakes  produced by these protoplanets
can be discerned even though the the medium is turbulent and
produces erratic perturbations of them. Thus the tidal perturbation
of the disc produces measurable effects.

Figure \ref{figbox2}  shows running time averages of the torques acting on the
protoplanet in simulations  Ba1--Ba4.
Apart from simulation Ba1, these  commence from when the protoplanets
were introduced into the simulations. In the case of Ba1, the running average
was started somewhat later.

The averages were calculated over time periods of typically
$50$ orbits at the centre of the box.
In each case the average torques from the regions exterior to and interior
to the protoplanet drag and accelerate the protoplanet as expected.
Although, as in the global simulations, fluctuations in the one sided
torques   can be very large amounting  to an order of magnitude or more
greater than the typical average value, the averages tend to approach
reasonably steady values. Note that in a shearing
box, symmetry considerations require that the
mean torques from the two sides of the disc
should ultimately be equal and opposite.
However, some noise remains 
even after fifty orbits.   The noise is stronger in the embedded
cases  and remains at the five to ten percent level
after fifty orbits. The resulting mean net torque which would
be zero in a laminar simulation suffers  corresponding fluctuations.
However, the effects of noise seems much less severe in simulations
Ba3 and Ba4 in which the disc is non linearly perturbed, in agreement with
the global run G5.
In these cases turbulent fluctuations seem to be less able
to modulate the torque estimates and non zero net torques 
arising from features added to bias the box are more readily measurable.

\begin{figure}
\centerline{
\epsfig{file=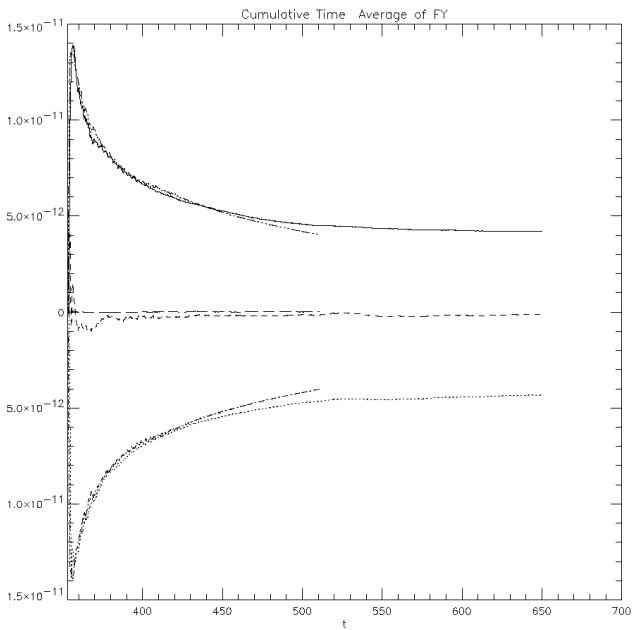,width=\columnwidth} }
\caption[]
{This figure shows running time averages of the torques acting on the
protoplanet in simulation Ba4 commencing from when the protoplanet
was introduced. The uppermost curve extending to larger times gives the
net torque acting on the planet from the inner regions while the corresponding
lowermost curve gives the contribution from the outer regions. The central short dashed
curve gives the net torque. For comparison results from a laminar  simulation  with the same
parameters but no magnetic field are presented. Corresponding plots from the two
simulations can be almost superposed but the laminar case terminates at an earlier time.}
\label{figbox3}
\end{figure}

To illustrate the reduced effects
of noise on the simulations with
strongly perturbing protoplanets, we
compare running time  averages of the torques acting on the
protoplanet in simulation  Ba4 with those
obtained from the corresponding laminar disc simulation
in figure \ref{figbox3}. The  averages  commence from when the protoplanet
was introduced. 
Corresponding plots from the two
simulations are very similar. Note that the torques tend to relatively weaken
at later times in the laminar case. This is because this simulation
with  $(M_p R^3/(M_* H^3) =2 $ produces a wider and 
deeper gap than the turbulent case (see paper III). This in turn  results in
less matter for the protoplanet to interact with locally on average
and hence weaker average torques at later times. 

Finally we comment on the  magnitudes of the time
averaged one sided torques. As for the global runs
these are similar for turbulent and corresponding laminar disc simulations.
For the shearing box simulations, the fiducial value  expected for the
one sided  linear torques  in  the 2D laminar case
with no potential softening  is 
\be  F_{y0} = 0.4\pi \Sigma R^3\Omega_p^2 \left({M_p^2R^3\over M_* H^3}\right ) ,\ee
where  the  surface density
$\Sigma = \langle \rho \rangle H ,$
with $\langle \rho \rangle$ being the volume and time averaged density
in the box (see Ward 1997). For simulation Ba1, $ F_{y0} = 10^{-13}$
in our computational units making the typical averaged torques
about fifty percent of the fiducial value. Such a reduction may
occur as  a result of softening the gravitational potential
due to the protoplanet (see section \ref{calibration} and paper III).
 This would indicate
that the magnitude of the torques is essentially given by the
laminar disc theory. However, the turbulence adds a significant
noise component when the protoplanet mass is small
enough to place the response in the linear regime.
We comment that the averaged torque magnitudes in Ba1 and Ba2 are
consistent with a scaling $\propto M_p^2$ appropriate
to the linear response regime but that the torque is
weaker than suggested by such a scaling in Ba3 and Ba4.
This is consistent with the response being non linear in those cases.

\section{Discussion} \label{conclusions}

In this paper we have performed both global and local simulations  
of embedded protoplanets interacting with a turbulent disc
and studied the behaviour of the torques exerted between
the disc and  protoplanets and the consequences for orbital migration.
The global simulations were for a disc with $H/r =0.07$
that exhibited MHD turbulence with zero net magnetic flux
with mean $\alpha \sim 0.007.$
The protoplanet masses considered were $M_p= 3, 10$ and $30$ Earth masses,
and 3 Jupiter masses.
The local shearing box simulations can be characterized
by values of the dimensionless parameter $M_p R^3/(M_* H^3).$  The simulations  adopted 
$0.1, 0.3, 1.0, $ and $2.0$ respectively. The first two of these are directly
comparable to the global simulations with $M_p= 10M_{\oplus},$ and $M_p = 30M_{\oplus}$
respectively. The latter two have  gap forming protoplanets, but whose masses
correspond to less massive planets than the 3 Jupiter mass planet considered in the
global run G5, and enable
the behaviour of the torques in the non linear gap forming regime
to be studied. For a first study, the protoplanets considered here were 
held in fixed circular orbit.

It was always found that the instantaneous  torque experienced by a protoplanet
was a highly variable quantity on account of the
protoplanet interacting with the turbulent density wakes that shear past it.
For low mass protoplanets that are not able to begin to
form a gap, the torque is dominated by
these fluctuations, such that  at any particular time,
the usual distinction between inner (positive)
and outer (negative)
disc torques is blurred. The net torque experienced by embedded
protoplanets oscillates between negative and positive values, such that
the protoplanet migration is likely to occur as a random walk.
This is in contrast to the monotonic inward drift normally associated with
type I migration. 

In order to average out the erratic behaviour
of the instantaneous torque a running time average is considered.
We considered contributions from the inner and outer disc.
We have been able to do this over a $20$ orbital period timescale.
Although these running averages took on values characteristically
expected for type I migration, large fluctuations remained
such that the net torques failed to converge to well defined
values over 
the run times that are currently
feasible. This is in contrast to laminar disc simulations
for which convergence is achieved on an orbital time scale.   

Fluctuations in the averaged one sided and net
torques were most severe for the smaller embedded masses.
Thus for the $3 M_{\oplus}$ global simulation, fluctuations
in the one sided torques  averaged over $\sim 20$ orbits were comparable
to the averaged torques themselves resulting in an indeterminate direction
of migration. However, in the case of the $30 M_{\oplus}$ global simulation, fluctuations
in the one sided averaged torques were relatively smaller at about $20$
percent of the mean values  resulting in a more clearly
delineated but still erratic inward migration.

The  existence of large fluctuations in the torques exerted
by the disc on embedded low mass protoplanets and
lack of knowledge of how these behave on long time scales makes definitive statements
about the direction and rate of migration  difficult to make.
However, even in spite of large instantaneous fluctuations of more than
one order of magnitude larger than the mean values, characteristic values 
of one sided torques
averaged over a $20$ orbit time scale are consistent with laminar type I estimates.
Thus, when this estimate gives inward migration to
the central object,  we might expect the fluctuations to result in increased survival prospects
for a subset of a statistical ensemble of embedded protoplanets.

The behaviour of the fluctuations   on long timescales  remains
an important  issue that  is impractical to explore
at the present time. If there are significant fluctuations generated in the global
simulations that occur on the longest evolutionary time scales, convergence
of torque running means becomes for practical purposes impossible to achieve and
the migrational behaviour of low mass protoplanets considered as an ensemble
would be very different from predictions of type I migration theory. 

We found that noise levels were relatively smaller in the local shearing box simulations.
This gives some support to the  idea of the
existence of global fluctuations with long characteristic times
in the global simulations which are expected to be absent in the box simulations.

We comment that the zero net magnetic flux models that we consider here 
generally give rise to turbulent discs that are more quiescent
than those which arise when the magnetic fields has net flux,
especially if a
net poloidal field is present (e.g. Hawley, Gammie, \& Balbus 1996;
Hawley 2001; Steinacker \& Papaloizou 2002).
Strong density fluctuations and contrasts, including persistent gaps,
may arise in such discs, and would have a profound impact on the
migration of embedded protoplanets. In particular, if MHD turbulence gives rise
to global disc structures that lead to varying migration rates as a function of 
position in the disc, then migration of embedded protoplanets into these
low--migration regions is likely to occur, increasing the lifetime of these
planets in the disc. 

The results  of both the global and local simulations show the 
same trend as a function
of planet mass. For very low mass protoplanets the turbulent density wakes
have higher amplitude than the spiral wake generated by the
protoplanet. The torques exerted on the protoplanet are  then such that
the instantaneous turbulent fluctuations are very much larger than the running means.
However, as the planet mass increases, the perturbation it makes on the disc
starts to dominate its local neighbourhood.
The expected separation between inner and outer disc torques becomes apparent
and fluctuations become smaller relative to the running mean
torques.
Eventually gap formation
occurs. At this point  we find a weakening  of the torques
exerted by the disc on the protoplanet due to the evacuation of gap material.
There is then a transition to type II
migration at a
rate determined by the angular momentum transport in the distant parts of the  disc
unaffected by the protoplanet. This has already been seen in  the simulation
of a $5$ Jupiter mass protoplanet described in paper II.

\subsection{Acknowledgements} 

The computations reported here were performed using the UK Astrophysical 
Fluids Facility (UKAFF).

{}
\end{document}